\documentclass[journal]{IEEEtran}
\makeatletter\if@twocolumn\PassOptionsToPackage{switch}{lineno}\else\fi\makeatother

\usepackage{graphicx}
\usepackage[T1]{fontenc}
\ifCLASSINFOpdf
\else
\fi
\usepackage{amsmath}
\usepackage[cmintegrals]{newtxmath}
\usepackage{url,multirow,morefloats,floatflt,cancel,tfrupee}
\makeatletter

\AtBeginDocument{\@ifpackageloaded{textcomp}{}{\usepackage{textcomp}}}
\makeatother
\usepackage{colortbl}
\usepackage{xcolor}
\usepackage{pifont}
\usepackage{threeparttable}
\usepackage[nointegrals]{wasysym}
\makeatletter

\def\mcWidth#1{\csname TY@F#1\endcsname+\tabcolsep}

\def\cAlignHack{\rightskip\@flushglue\leftskip\@flushglue\parindent\z@\parfillskip\z@skip}
\def\rAlignHack{\rightskip\z@skip\leftskip\@flushglue \parindent\z@\parfillskip\z@skip}

\@ifundefined{etal}{}{}

\usepackage{ifxetex}
\ifxetex\else\if@twocolumn\@ifpackageloaded{stfloats}{}{\usepackage{dblfloatfix}}\fi\fi

\AtBeginDocument{
\expandafter\ifx\csname eqalign\endcsname\relax
\def\eqalign#1{\null\vcenter{\def\\{\cr}\openup\jot\m@th
  \ialign{\strut$\displaystyle{##}$\hfil&$\displaystyle{{}##}$\hfil
      \crcr#1\crcr}}\,}
\fi
}

\AtBeginDocument{%
  \@ifpackageloaded{endfloat}%
   {\renewcommand\efloat@iwrite[1]{\immediate\expandafter\protected@write\csname efloat@post#1\endcsname{}}}{\newif\ifefloat@tables}%
}%

\def\BreakURLText#1{\@tfor\brk@tempa:=#1\do{\brk@tempa\hskip0pt}}
\let\lt=<
\let\gt=>
\def\processVert{\ifmmode|\else\textbar\fi}

\@ifundefined{subparagraph}{
\def\subparagraph{\@startsection{paragraph}{5}{2\parindent}{0ex plus 0.1ex minus 0.1ex}%
{0ex}{\normalfont\small\itshape}}%
}{}

\newcommand\role[1]{\unskip}
\newcommand\aucollab[1]{\unskip}
  
\@ifundefined{tsGraphicsScaleX}{\gdef\tsGraphicsScaleX{1}}{}
\@ifundefined{tsGraphicsScaleY}{\gdef\tsGraphicsScaleY{.9}}{}
\def\checkGraphicsWidth{\ifdim\Gin@nat@width>\linewidth
	\tsGraphicsScaleX\linewidth\else\Gin@nat@width\fi}

\def\checkGraphicsHeight{\ifdim\Gin@nat@height>.9\textheight
	\tsGraphicsScaleY\textheight\else\Gin@nat@height\fi}

\def\fixFloatSize#1{}
\let\ts@includegraphics\includegraphics

\def\inlinegraphic[#1]#2{{\edef\@tempa{#1}\edef\baseline@shift{\ifx\@tempa\@empty0\else#1\fi}\edef\tempZ{\the\numexpr(\numexpr(\baseline@shift*\f@size/100))}\protect\raisebox{\tempZ pt}{\ts@includegraphics{#2}}}}

\AtBeginDocument{\def\includegraphics{\@ifnextchar[{\ts@includegraphics}{\ts@includegraphics[width=\checkGraphicsWidth,height=\checkGraphicsHeight,keepaspectratio]}}}

\DeclareMathAlphabet{\mathpzc}{OT1}{pzc}{m}{it}

\def\URL#1#2{\@ifundefined{href}{#2}{\href{#1}{#2}}}

\def\UrlOrds{\do\*\do\-\do\~\do\'\do\"\do\-}%
\g@addto@macro{\UrlBreaks}{\UrlOrds}

\edef\fntEncoding{\f@encoding}

\makeatother

\newif\ifmultipleabstract\multipleabstractfalse%
\usepackage{tabulary}
\makeatletter
\AtBeginDocument{\@ifpackageloaded{longtable}{%
\def\LT@makecaption#1#2#3{%
  \LT@mcol\LT@cols c{\hbox to\z@{\hss\parbox[t]\LTcapwidth{%
    \sbox\@tempboxa{#1{#2: } #3}%
    \ifdim\wd\@tempboxa>\hsize
      #1{#2: }\textsc{#3}%
    \else
      \hbox to\hsize{\hfil\box\@tempboxa\hfil}%
    \fi
    \endgraf\vskip\baselineskip}%
  \hss}}}
}{}}
\makeatother

   \makeatletter
  \def\fig@textbf{\textbf}
   \AtBeginDocument{\renewcommand\floatc@plain[2]{\setbox\@tempboxa\hbox{{\footnotesize#1.}\footnotesize\hskip.5em#2}%
    \ifdim\wd\@tempboxa>\hsize {\fig@textbf{\footnotesize#1.}}\footnotesize\hskip.5em#2\par
        \else\hbox to\hsize{\hfil\box\@tempboxa\hfil}\fi}}
    \makeatother
  
\usepackage{float}

\begin{document}

%

        \title{SliT: A strip-sensor readout chip\\
        with subnanosecond time walk\\
        for the J-PARC muon $g-2$/EDM experiment}
      
\author{
        Tetsuichi~Kishishita,
        Yutaro~Sato,
        Yoichi~Fujita,
        Eitaro~Hamada,
        Tsutomu~Mibe,
        Tsubasa~Nagasawa,
        Shohei~Shirabe,
        Masayoshi~Shoji,
        Taikan~Suehara,
        Manobu~M.~Tanaka,
        Junji~Tojo,
        Yuki~Tsutumi,
        Takashi~Yamanaka,
        Tamaki~Yoshioka,
        \thanks{Tetsuichi~Kishishita, Yutaro~Sato, Yoichi~Fujita, Eitaro~Hamada, Tsutomu~Mibe, Masayoshi~Shoji, Manobu~M.~Tanaka are with Institute of Particle and Nuclear Studies, KEK, High Energy Accelerator Research Organization, 1-1 Oho, Tsukuba, 3050801, Ibaraki, Japan, Tel.:~+81-29-864-5384}
        \thanks{Tetsuichi~Kishishita is with Department of Accelerator Science, SOKENDAI (The Graduate University for Advanced Studies), e-mail: kisisita@post.kek.jp (Corresponding author).}
        \thanks{Yutaro~Sato is with Graduate School of Science and Engineering, Ibaraki University, 2-1-1 Bunkyo, Mito, 3108512, Ibaraki, Japan}
        \thanks{Takashi~Yamanaka is with Faculty of Arts and Science, Kyushu University, 744 Motooka, Nishi-ku, 8190395, Fukuoka, Japan}
        \thanks{Tsubasa~Nagasawa, Shohei~Shirabe, Taikan~Suehara, Junji~Tojo, Yuki~Tsutsumi are with Department of Physics, Kyushu University, 744 Motooka, Nishi-ku, 8190395, Fukuoka, Japan}
        \thanks{Tamaki~Yoshioka is with Research Center for Advanced Particle Physics, Kyushu University, 744 Motooka, Nishi-ku, 8190395, Fukuoka, Japan}
        }

\maketitle 

\begin{abstract}
A new silicon-strip readout chip named ``SliT'' has been developed for the 
measurement of the muon anomalous magnetic moment and electric dipole moment 
at J-PARC. The SliT chip is designed in the Silterra 180~nm CMOS technology with 
mixed-signal integrated circuits. An analog circuit incorporates a 
conventional charge-sensitive amplifier, shaping amplifiers, and two distinct 
discriminators for each of the 128~identical channels. A digital part includes 
storage memories, an event building block, a serializer, and LVDS drivers.
A distinct feature of the SliT is utilization of the zero-crossing architecture, 
which consists of a CR-RC filter followed by a CR circuit as a voltage 
differentiator. This architecture allows generating hit signals with 
subnanosecond amplitude-independent time walk, which is the primary 
requirement for the experiment. The test results show a time walk of 
$\boldsymbol{0.38 \pm 0.16}$~ns between 0.5 and 3~MIP signals.
The equivalent noise charge is $\boldsymbol{1547 \pm 75~e^{-}}$~(rms) at 
$\boldsymbol{C_{\rm det} =33}$~pF as a strip-sensor capacitance.
The SliT128C satisfies all requirements of the J-PARC muon $\boldsymbol{g-2}$/EDM experiment.
\end{abstract}


%
\IEEEpeerreviewmaketitle

\section{Introduction }
The anomalous magnetic moment ($g-2$) and the electric dipole moment (EDM) of 
the muon are sensitive probes of new physics beyond the Standard Model (SM).
The SM is a well-tested physics theory,
however, there exists a discrepancy in the muon $g-2$ between the 
SM prediction~\cite{PhysRevD.101.014029,Davier:2019can} and its most precise 
measurement by the E821 collaboration at Brookhaven National Laboratory (BNL) 
by more than three standard deviations~\cite{Bennett:2006fi}.
An experiment to improve the precision of the muon $g-2$ is ongoing at 
Fermilab which uses the upgraded apparatus and plans to
increase statistics~\cite{Grange:2015fou}.
On the other hand, another measurement of the muon $g-2$/EDM is currently under 
preparation based on a different experimental approach to make a definitive 
conclusion. This experiment applies an innovative technique using a 
reaccelerated thermal muon beam at the Japan Proton Accelerator Research 
Complex (J-PARC)~\cite{Abe:2019thb}.

A primary goal of the experiment is to measure the muon $g-2$ with a 
precision of 0.1~parts per million (ppm), which corresponds to an improvement 
by a factor of five compared to the BNL experiment.
At the same time, we seek for the muon EDM with a sensitivity of 
$10^{-21}$~$e\cdot$cm.
The reaccelerated thermal muon beam with a repetition rate of 25~Hz to be 
used at J-PARC is produced from a thermal muonium source,
followed by laser ionization and acceleration.
The reaccelerated muons with a momentum of 300~MeV/$c$ are stored under a 
3~T magnetic field.
The muon makes the orbital cyclotron motion in the mid-plane of the muon 
storage magnet.
A positron tracking detector is placed in the innner region of the muon 
storage ring for the detection of positrons from the muon decay, i.e., 
$\mu^{+}\rightarrow e^+\nu_e \bar{\nu}_{\mu}$.
Fig.~\ref{fig:g-2} shows the concept of the muon storage ring and positron 
detector in the experiment.
The 40~sensor vanes are aligned radially in the detection volume.
Each vane consists of single-sided p-on-n type silicon-strip 
sensors~\cite{HPKS13804} with mutually orthogonal strips on both sides.
A two-dimensional position of a positron track is therefore determined by 
two layers of the strip sensors.
Circular positron tracks are reconstructed by connecting two-dimensional positions and their momenta are derived from the curvature.
To maximize the statistical sensitivity of the muon $g-2$,
we will use the positron tracks with a momentum in the range from 200~MeV/$c$ to 275~MeV/$c$ for the analysis.
The measurement will be performed in an interval following a fill of 33~$\upmu$s,
which is five times longer than the lifetime of the muon with the momentum 
of 300~MeV/$c$.

An integrated circuit with a fast response and high granularity is used to 
readout data from the silicon strip detectors.
The integrated circuit is equipped with a deep buffer memory to store the 
data for a beam spill.
The stored data is readout before the next beam bunch coming.
To meet the experimental requirements, a series of custom front-end readout 
chips named ``SliT'' has been developed in 180~nm CMOS
technology. 
We have newly developed a full channel chip named ``SliT128C'', and its 
performance has been tested in the laboratory.
In this paper, we describe the ASIC~(application-specific integrated circuit)
design in section~II, report the evaluation setup and the performance test 
results in section~III, and finally give conclusions in section~IV.

\bgroup
\fixFloatSize{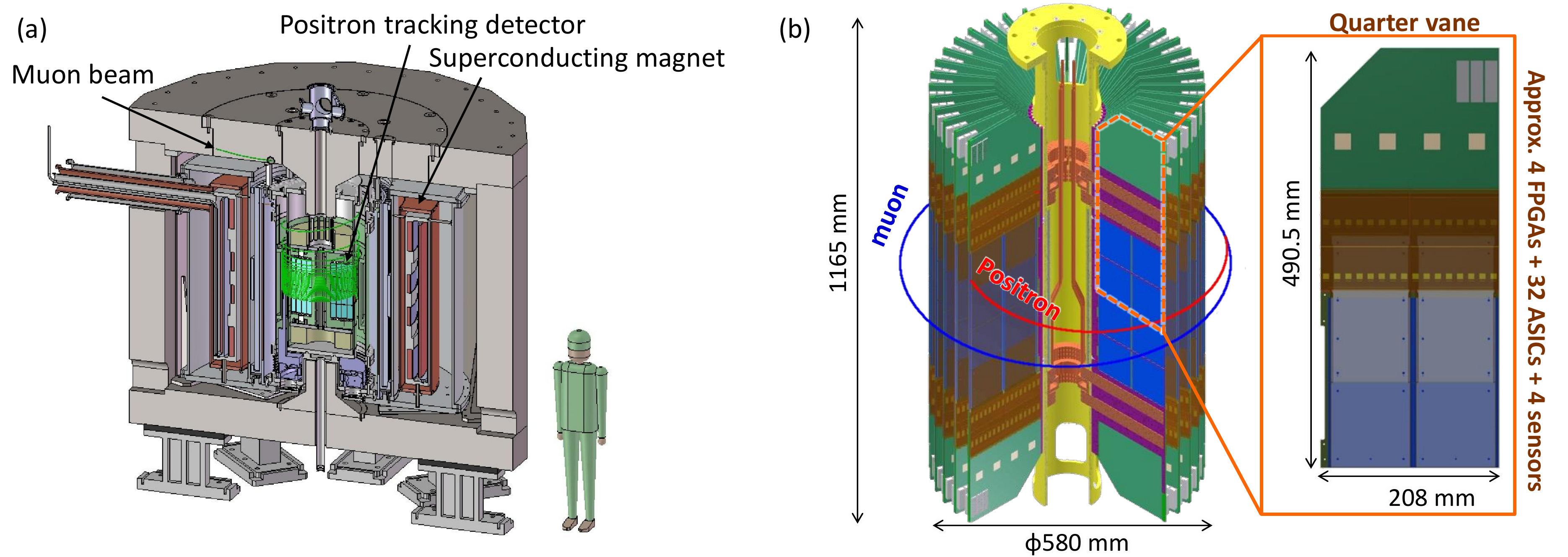}
\begin{figure*}[!htbp]
\centering \makeatletter\IfFileExists{magnet_detector_vane.pdf}{\includegraphics{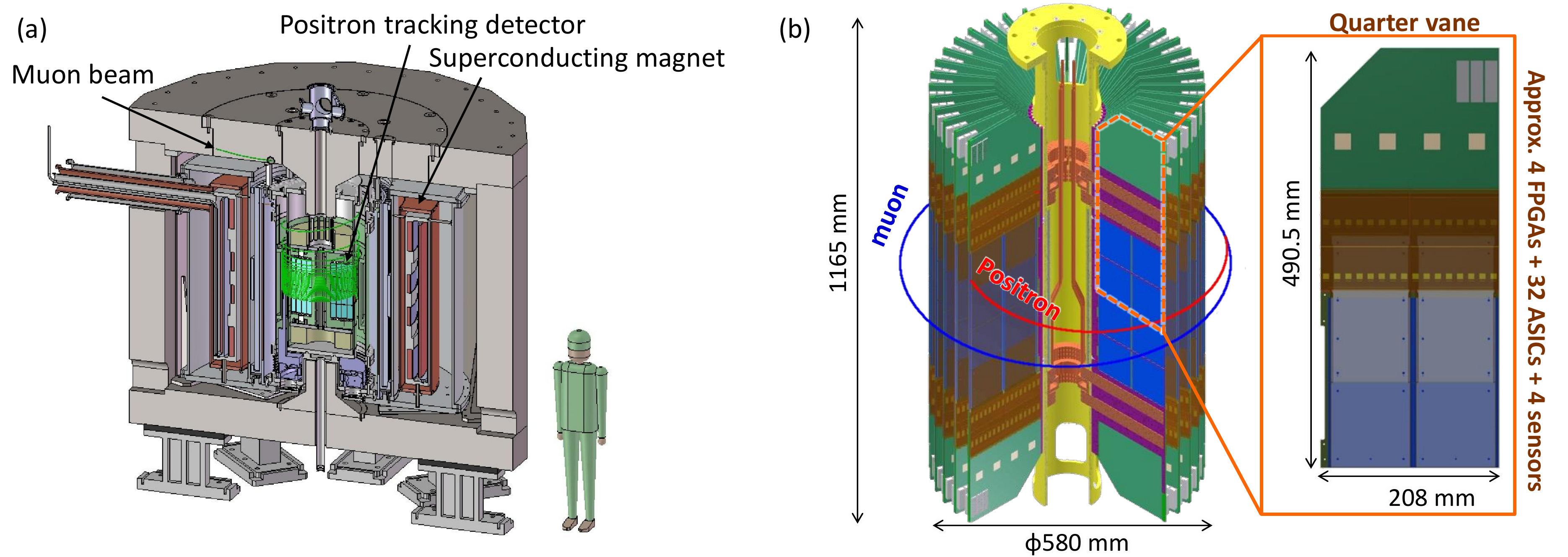}}{}
\makeatother 
\caption{{(a) Muon storage ring and positron tracking detector in the muon 
$g-2$/EDM experiment at J-PARC. (b) Perspective view of silicon-strip 
detectors and a closeup view of the so-called quarter vane, which consists 
of four sensors and 32 front-end ASICs.}}
\label{fig:g-2}
\end{figure*}
\egroup

\section{Architecture of the ASIC}

\subsection{Experimental requirements and developmental background}
The readout ASIC was implemented in the Silterra 180~nm CMOS technology. 
Table~\ref{tab:technology} summarizes the technological parameters.
The front-end ASICs for silicon-strip sensors are requested to tolerate a 
high hit rate up to 1.4~MHz per strip with a stability to dynamically changing 
hit rates by a factor of 150 from the beginning to the end of measurements.
In order to minimize the pileups, a pulse width for a typical minimum-ionizing 
particle~(MIP) signal has to be smaller than 100~ns. Here, the pulse width is 
defined from a 1~MIP output, which intersects a threshold set at 0.3~MIP. 
For the further reduction of the pileups and precise detection of circular positron tracks from the muon decays,
a subnanosecond time walk is of critical importance in analog processing blocks of the readout ASIC.
The experimental requirements translated to concrete circuit specifications 
are summarized in Table~\ref{tab:requirement}.

AC-coupled single-sided p-on-n type silicon-strip sensors with a double-metal 
structure were made by Hamamatsu Photonics K.K.~\cite{HPKS13804}.
The sensor thickness is 320~$\upmu$m.
In this paper, the most probable value of the MIP charge is defined as 
3.84~fC, which corresponds to an input charge of about 24,000~$e^{-}$.
The sensor capacitance is approximately 17~pF, and the total detector capacitance $C_{\rm det}$ seen from an input of the ASIC is estimated to be less than 30~pF. This value includes the parasitic capacitance coming from Flexible Printed Circuits (FPCs)~\cite{Yamanaka}, which connect strip sensors with ASICs.
Since the maximum wiring length inside the FPCs is about 250~mm,
the parasitic capacitance from the FPCs is not negligible.
The equivalent noise charge (ENC) of less than 1600~e$^-$ at $C_{\rm det}=30$~pF 
is required.

In the previous studies, we started from a small-scale analog prototype in 
Silterra 180~nm CMOS technology.
Then we fabricated module-prototype chips named ``SliT128A''~\cite{Aoyagi:2019ujx} and ``SliT128B''~\cite{HSTD12_proceeding}.
The module-prototype ASICs included 128 readout channels, buffer memories, 
and other digital processing circuits to store and cope with timing 
information from the strip sensors. Since the time walk performance was 
measured as $\sim$17.2~ns in SliT128A, we introduced an additional CR block 
after the semi-Gaussian band-pass filter in SliT128B (see the next section for detailed description of this measure). As a result, 
the time walk improved to $\sim$2~ns, however, it was still larger than the 
requirement. A new readout chip named ``SliT128C'' improves the overall analog 
performance, e.g., the ENC, time walk, and jitter, by optimizing transistor parameters of bias circuits in SliT128B.

Fig.~\ref{fig:slit128c_photo} shows the photograph of SliT128C. Inputs of 128 channels
are aligned in staggered pads with a pitch of 100~$\upmu$m. The incident signals are processed from the left analog block to the right digital block. Power
supply is provided from the top and bottom pads only, and digital outputs 
are accessed on the right-hand side.
To suppress the digital cross talks, the analog and digital blocks are 
enclosed in distinct deep N-well islands.

\begin{table}[!htbp]
\caption{{Technological parameters} }
\label{tab:technology}
\def\arraystretch{1}
\ignorespaces 
\centering 
\begin{tabulary}{\linewidth}{p{\dimexpr.50\linewidth-2\tabcolsep}p{\dimexpr.50\linewidth-2\tabcolsep}}
\hline \hline
Technology &  Silterra 180~nm CMOS, 6 metals\\
Process options & Deep NWell, MIM cap. (4~fF/$\upmu$m$^2$) \\
& high resistive p+ poly (1050~$\Omega$/sq.) \\
Chip size  &  6.54~mm $\times$ 7.2~mm\\
Thickness & 381~$\upmu$m \\
Supply rail & 1.8~V (core/IO)\\
\hline
\end{tabulary}\par 
\end{table}

\bgroup
\begin{figure}[!htbp]
\centering \makeatletter\IfFileExists{slit128c_photo_crop.jpg}{\includegraphics[width=7cm]{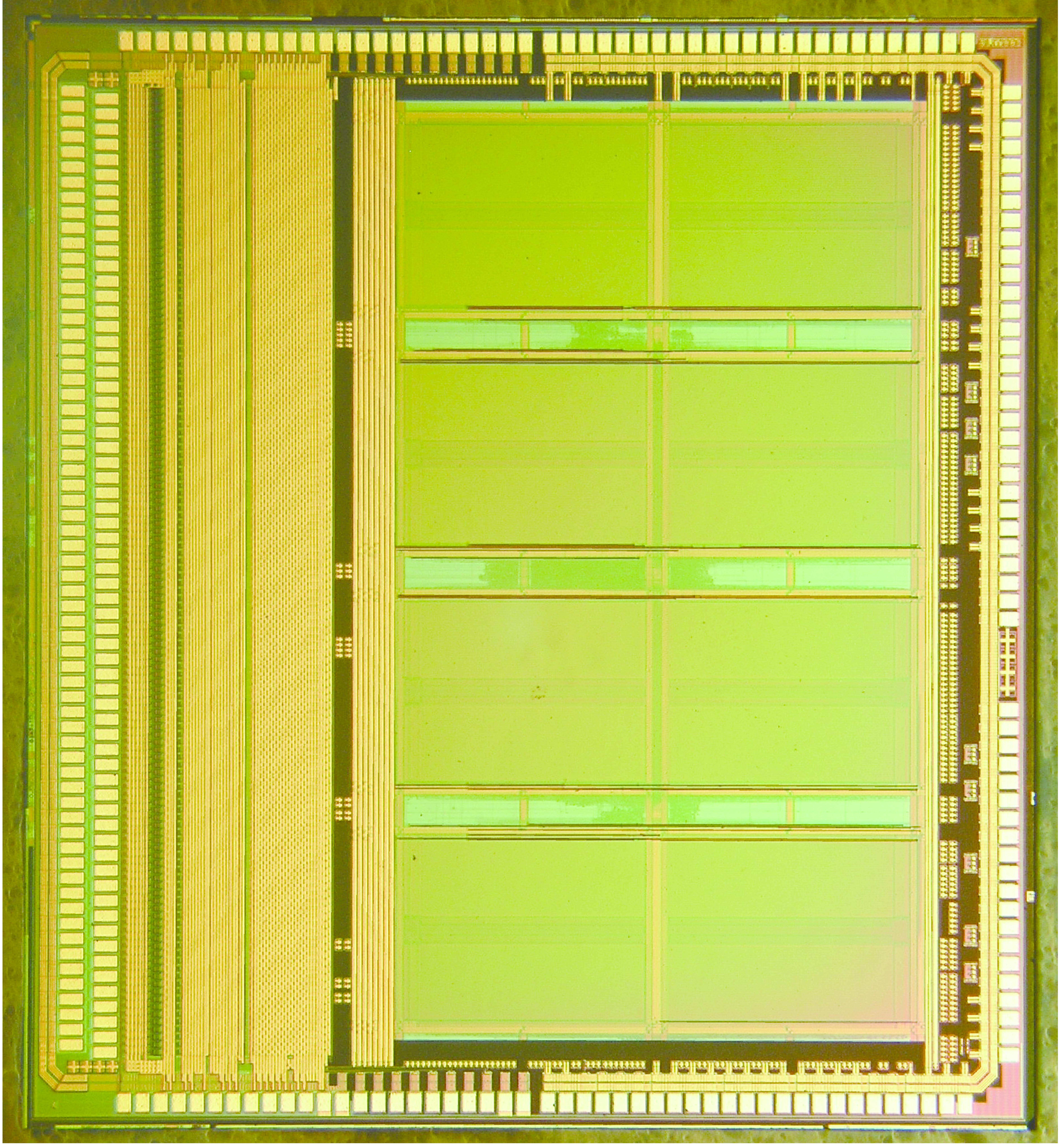}}{}
\makeatother 
\caption{{Photograph of the ASIC fabricated in the Silterra 180~nm CMOS technology.
The chip size is 6.54~mm $\times$ 7.2~mm with a memory area density of 40\%.}}
\label{fig:slit128c_photo}
\end{figure}
\egroup

\subsection{Analog processing block}
Fig.~\ref{fig:analog} shows the signal processing chain of each channel. 
The strip sensor is AC-coupled to an input of the charge-sensitive amplifier 
(CSA), while test pulses can be injected via an AC-coupling capacitor of 
100~fF. The CSA is based on a folded cascade configuration with a p-channel 
input transistor ($W/L=8~\upmu$m$/180$~nm). In terms of noise performance, we 
increased the folding number of the input transistor to $M~=~250$, which are 
placed in parallel. The feedback capacitor $C_{\rm f}$ was implemented by a 
metal-insulator-metal structure with a value of 170~fF. A transfer gate 
type FET was employed for the CSA DC-feedback component. 

The CSA output is fed into a CR-RC shaping amplifier composed of a 
pole-zero cancellation circuit (PZC) and a second-order integration low-pass 
filter. The capacitors and resistance values in the PZC were selected to meet 
the equation of $C_{\rm f}\cdot R_{\rm f}=C_{\rm pz}\cdot R_{\rm pz}$. In the 
semi-Gaussian shaper, high-resistance p+ poly-silicons were chosen as 
resistors. Strictly speaking, the parameters of $C_{\rm sh1/sh2}$ and $R_{\rm sh1/sh2}$ have to meet the equation of $C_{\rm sh1}\cdot R_{\rm sh1}=4C_{\rm sh2}\cdot R_{\rm sh2}$ for the ideal case that the network functions as the second-order CR-RC 
shaper \cite{kisisita}, however, we have tuned the actual values to minimize 
time walks by a SPICE simulation. 

The shaper output is amplified by an inverting amplifier by a factor of two. 
This function is selectable with a CMOS switch on the outside chance that 
voltage slopes of the shaper output affect timing accuracy. After the inverting amplifier, the output is fed into two signal paths, one of which is buffered 
and connected to a discriminator.
Baseline tuning is required to adjust the 
thresholds that is performed by two sets of 7-bit current DACs 
and resistors.
The other signal path is connected to the additional CR circuit, which functions as a voltage differentiator.


In the previous chip, SliT128A, the hit timing was produced from the 
first-order CR-RC shaper with a peaking time of 50~ns and a single 
discriminator. This method, however, was not able to reduce the time walk 
of the discriminator rising edge below the requirement of 1~ns, even after 
the fine tuning of the threshold. By implementing the differentiator after 
the CR-RC shaper, we improved the time walks between 0.5 and 3~MIP signals. 
By differentiating the semi-Gaussian shape, the output shape becomes bipolar, 
which traverses the baseline given by $V_{\rm base}$ in Fig.~\ref{fig:analog}.
The traversing timing becomes independent of input charges. This method corresponds to the 
so-called ``zero-crossing'' architecture and the traversing timing locates at a 
peaking time of the preceding CR-RC filter. As a result, the discriminator 
rising edge from the differentiator path remains less than $100$~ns, and at 
the same time, the time walk can be considerably improved. 

 The noise level after the voltage differentiator becomes worse than the ENC of the CR-RC shaper. 
To suppress triggers on noise, we preserved the CR-RC shaper path and used 
both information to define the final pulses by making an AND operation. The 
leading-edge timing of the final output is determined from the differentiator, 
while the trailing-edge timing is determined from the CR-RC shaper. 
Precisely speaking, the noise bandwidths of the CR-RC and CR-RC-CR are theoretically the same, i.e., the differentiator introduces a low-frequency cutoff, whereas the upper cutoff frequency remains about the same \cite{spieler}.
The noise level is related with the reduction in peak amplitude of the bipolar shapes. This degraded slope also produces larger time jitter (not the time walk itself), compared with the CR-RC filter. 
The time jitter or temperature effect on the time walk is, however, averaged due to large event statistics and not a severe problem in our experiment. The time walk and time jitter estimated by the Monte Carlo simulation, including the device mismatch and parasitic effect, are listed in Table~\ref{tab:requirement}.

The 
monitor lines were also implemented to examine analog waveforms at each 
processing point, e.g., the CSA, CR-RC shaper, differentiator etc. 
The 20-bit control registers are implemented to enable switches, i.e., a 
test pulse injection, the inverting amplifier, distinct discriminators, 
monitor lines, and tuning DACs for each channel.

\bgroup
\fixFloatSize{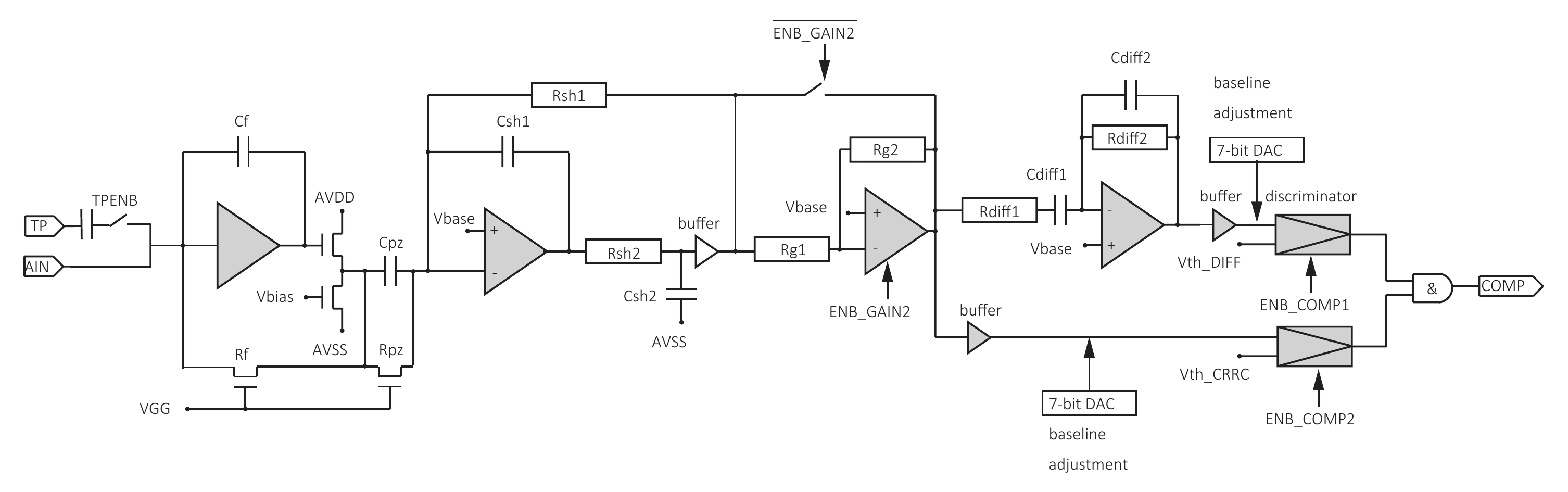}
\begin{figure*}[!htbp]
\centering \makeatletter\IfFileExists{fig3_rev200630.pdf}{\includegraphics{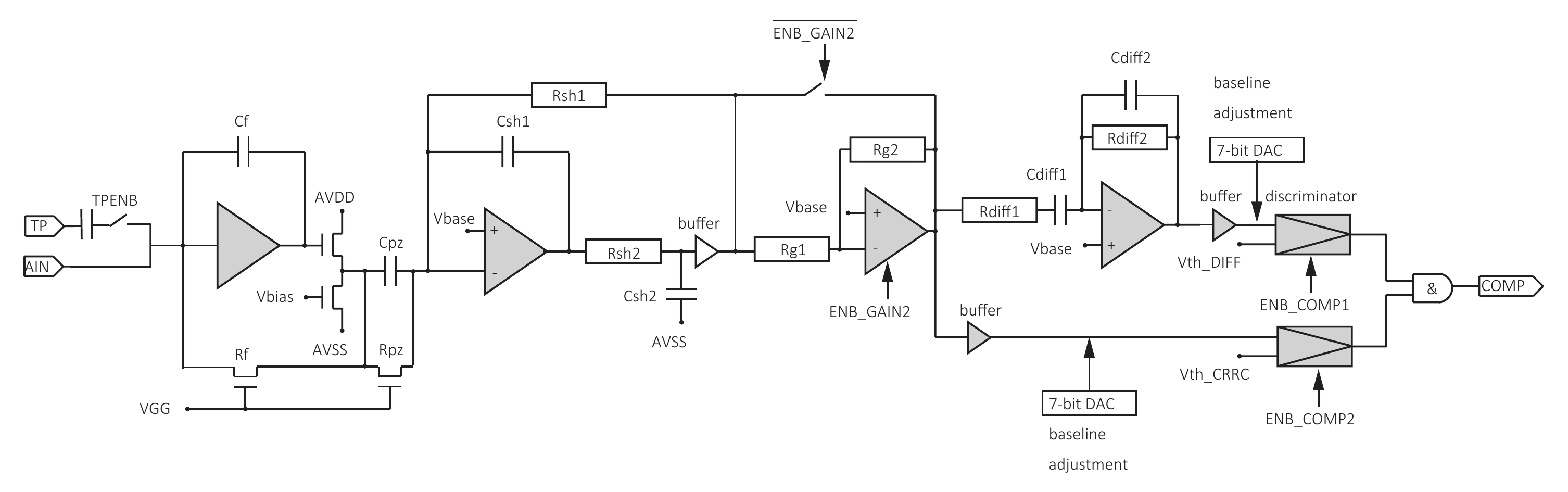}}{}
\makeatother 
\caption{{Signal processing chain for each channel. The capacitance values 
are $C_{\rm f}=170$~fF, $C_{\rm pz}=8\times C_{\rm f}$, $C_{\rm sh1}=154$~fF, $C_{\rm sh2}=40.5$~fF, $C_{\rm diff1}=54.5$~fF, $C_{\rm diff2}=5\times C_{\rm diff1}$. The resistance values are $R_{\rm sh1}=255~{\rm k}\Omega$, $R_{\rm sh2}=85~{\rm k}\Omega$, $R_{\rm g1}=100~{\rm k}\Omega$, $R_{\rm g2}=2\times R_{\rm g1}$, $R_{\rm diff1}=70~{\rm k}\Omega$, $R_{\rm diff2}=5\times R_{\rm diff1}$. $R_{\rm f}$ and $R_{\rm pz}$ are within the range of several M$\Omega$ adjusted by the gate voltage $V_{\rm GG}$.}}
\label{fig:analog}
\end{figure*}
\egroup

\subsection{Digital processing block}
Fig.~\ref{fig:dig} (a) shows the block diagram of the digital part. It 
consists of a signal interface (I/F), a memory controller, a serializer, 
a timing generator, and a parameter controller.
The timing generator block provides the ``Write Start'' and ``Read Start'' 
signals to the memory controller, which are synchronized with an external 
200~MHz clock.
The parameter controller block is a slow control circuit for the registers 
of both the analog and digital parts.
The signal interface block receives the discriminator outputs from the 
analog part. D-type flip-flops sample these signals by the external 200~MHz 
clock. After the first flip-flops, subsequent two D-type flip-flops resample these signals by 100~MHz clocks which are 180~degrees out of phase. This is 
because the sampling frequency of 200~MHz is close to the maximum clock 
frequency of logic circuits in the Silterra technology.

Fig.~\ref{fig:dig}(b) shows the block diagram of the memory controller. 
The memory controller block receives two kinds of data, which are 
synchronized by different 100~MHz clocks, from the signal I/F block. 
When the memory-write controller block receives the ``Write Start'' signal, 
it starts to store  data in SRAMs.
The memories 
have 8192~word depths for each channel, and the data can be stored within 
40.96~$\upmu$s ($=8192\times5$~ns). The storing process continues until the SRAMs 
are full. When the memory-read controller block receives the ``Read Start'' 
signal, it reads out all of the data from memories and sends them to the 
serializer block. The data is then sent out by the LVDS
drivers with a 50~MHz clock. The data are synchronized by the 
same 50~MHz clock in the backend electronics, e.g., an FPGA board in our 
applications, with the DDR~(Double Data Rate)
mode. 



\bgroup
\fixFloatSize{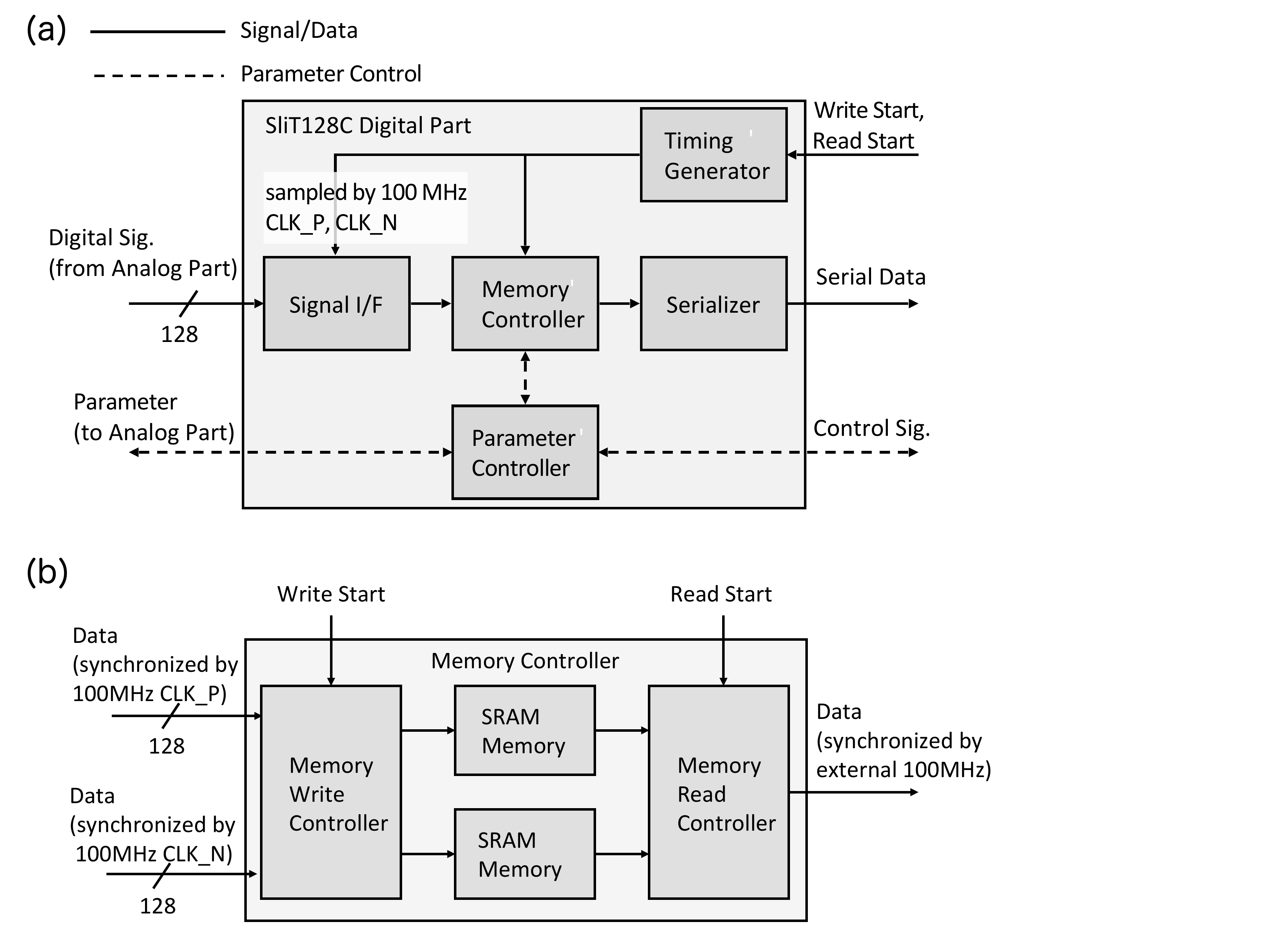}
\begin{figure}[!htbp]
\centering \makeatletter\IfFileExists{fig_dig1_rev200519.pdf}{\includegraphics[width=11cm]{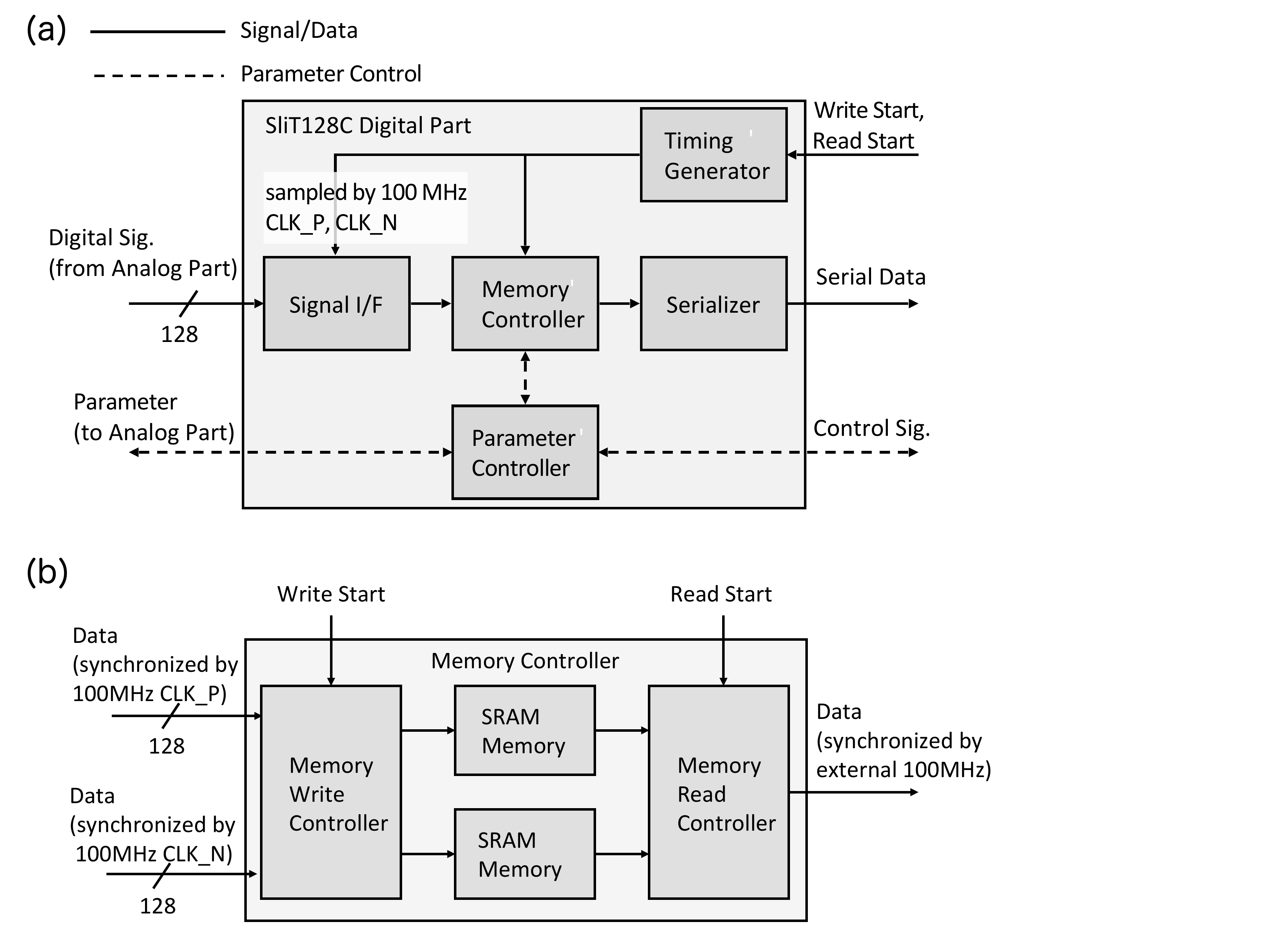}}{}
\makeatother 
\caption{{Block diagrams of (a) the digital part and (b) the memory controller.}}
\label{fig:dig}
\end{figure}
\egroup



\section{Performance Tests}

\subsection{Experimental setup}
The performance of the SliT128C was evaluated with a test pulse from a 
function generator.
A dedicated printed circuit board~(PCB) was designed for performance testing. 
Fig.~\ref{fig:evaltest_setup} shows the experimental setup.
A bare die is directly mounted on the PCB and a light-shielding box is 
placed on the chip.
The PCB is electrically connected to the ASIC by 25~$\upmu$m-diameter aluminum wires,
and also connected with a commercial field programmable gate array~(FPGA) board.
We chose the Digilent Nexys Video \cite{nexys} as an interface with a computer,
which is a commercial rigid PCB with a Xilinx Artix-7 FPGA~(XC7A200T).
This FPGA board provides an easy-to-use programming environment and allows 
easy modifications to test logics.
To store all data from SliT128C, a FIFO is programmed in the FPGA.
The stored data are transferred to a computer via an Ethernet cable with the 
SiTCP protocol~\cite{SiTCP}.
We evaluated the analog performance of 32 out of 128~readout channels,
each of which was equipped with 33~pF as a pseudo-strip-sensor capacitance 
for the test.

\bgroup
\fixFloatSize{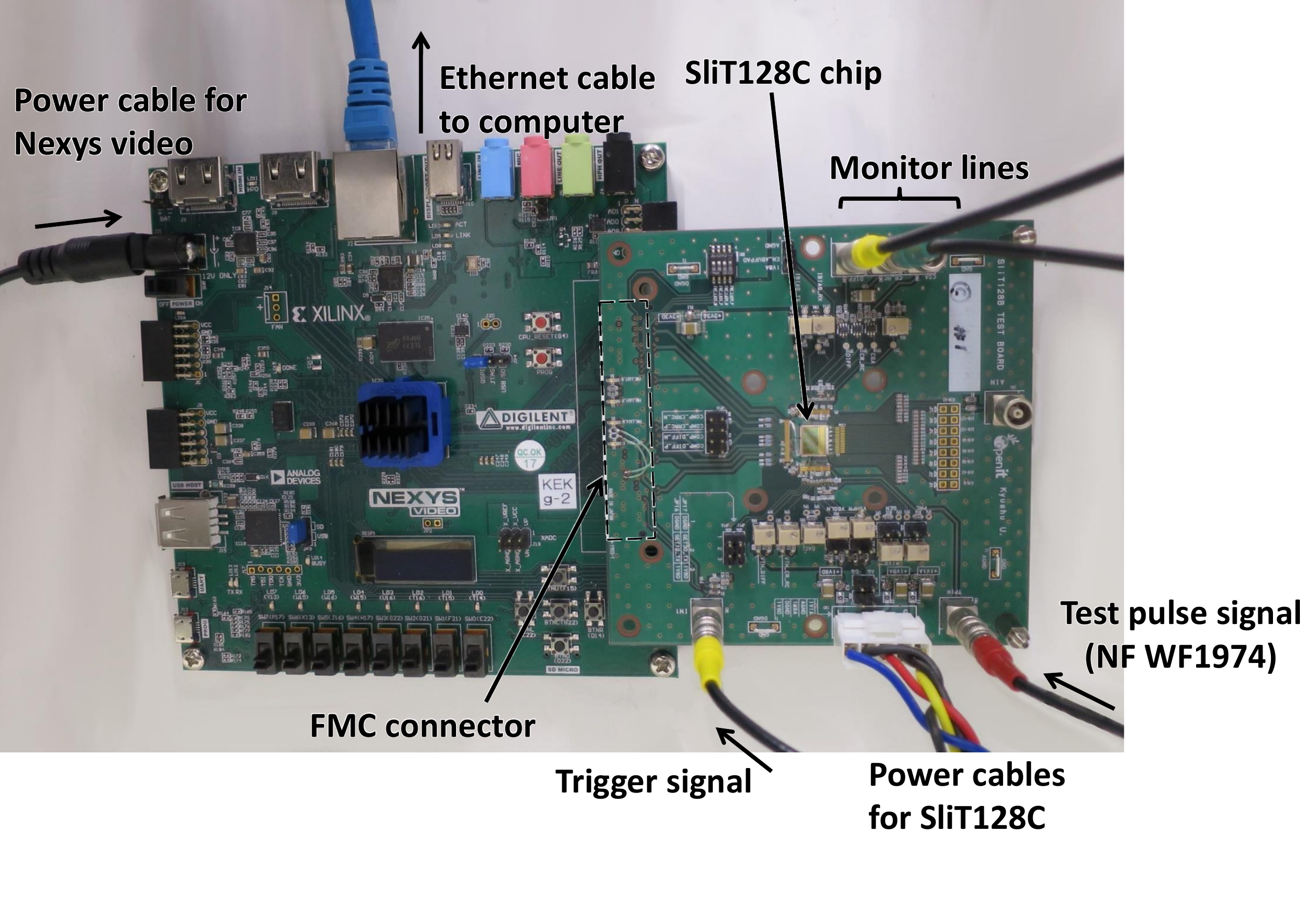}
\begin{figure}[!htbp]
\centering \makeatletter\IfFileExists{slit128c_setup_pic.pdf}{\includegraphics{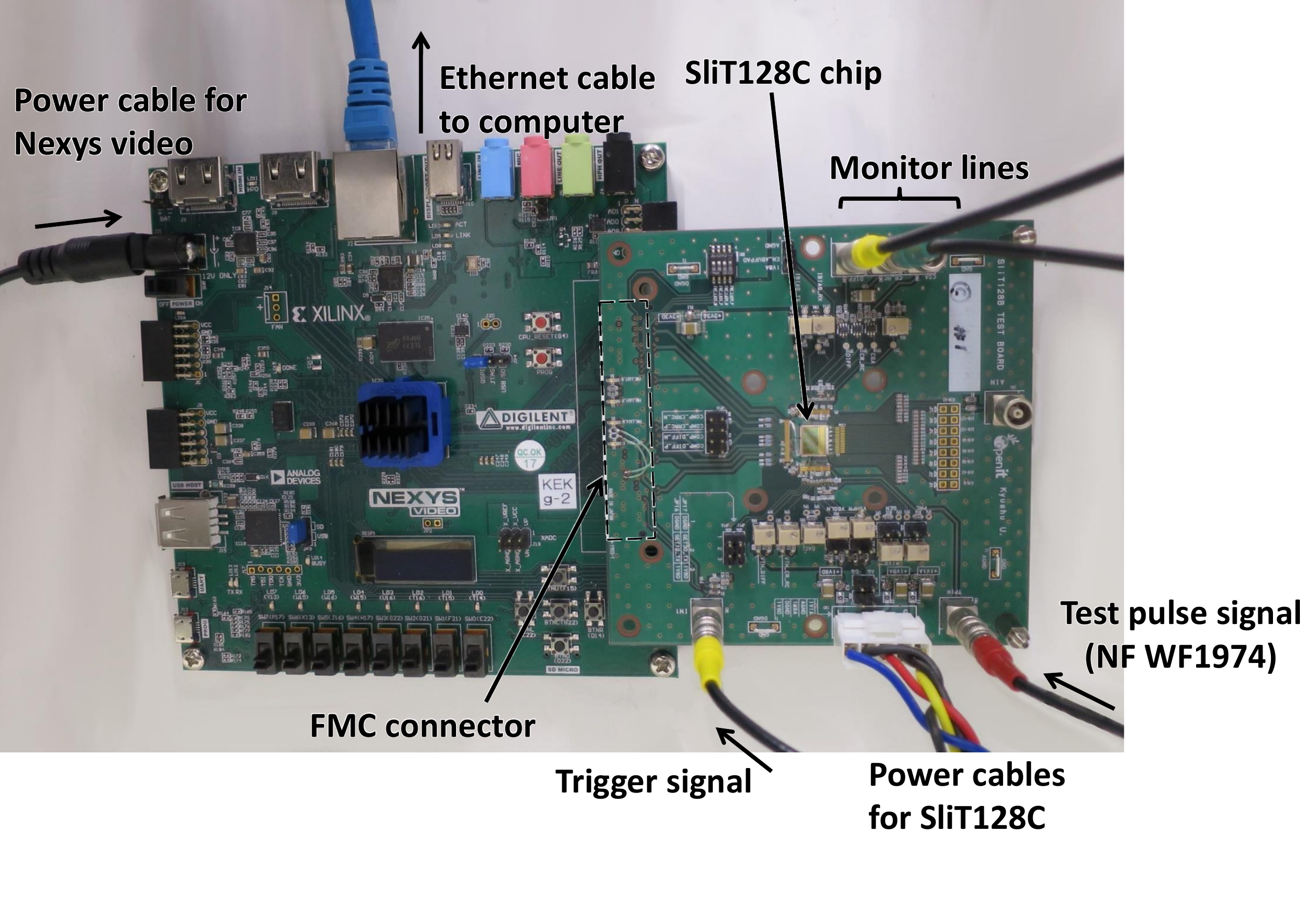}}{}
\makeatother 
\caption{{Setup for tests  of SliT128C performance.
A bare die is mounted on the dedicated PCB~(right side), which is connected 
with a commercial FPGA board of the Nexys Video~(left side) via an FPGA Mezzanine Card~(FMC) 
connector.
}}
\label{fig:evaltest_setup}
\end{figure}
\egroup

\subsection{Analog waveform and dynamic range}
The response of the monitor outputs was investigated to evaluate the analog 
waveform and dynamic range.
Fig.~\ref{fig:analog_monitor} shows typical analog output signals, when a 
test pulse with 1~MIP charge was injected. The outputs of discriminators are 
superimposed in black.
All monitor lines are functioning as designed.
The peaking time and pulse width were 64.2~ns and 99.2~ns, respectively.
Comparing with the values from a SPICE simulation~(35.1~ns and 75~ns) as 
shown in Table~\ref{tab:requirement},
we observe larger values due to a finite output-buffer performance to drive a 
large capacitive load of the bonding wire, $\sim$5~cm microstrips on the PCB, 
and the probe capacitance of $\sim$3.9~pF.
The actual pulse width at the discriminator must be shorter than the monitor 
values, and can be estimated only from the discriminator pulse width described 
in the following section.

Fig.~\ref{fig:dynamic_range} shows a pulse height of the CR-RC output as a 
function of injected charges. The wide dynamic range is needed to avoid the 
saturation for the high hit-rate environment. The integrated non-linearity~(INL) was 
measured as less than 1\% and 3.8\% in the ranges from 1.92~fC to 20~fC and 
from 1.92~fC to 30~fC, respectively.
An INL of 5\% is sufficient for our application.
The corresponding dynamic range is larger than 7.8~MIP.
We also confirmed that the channel-to-channel variation of offset voltages is 
within the amplitude of a 1.5~MIP charge that is reasonable to be compensated 
by 7-bit DACs.

\bgroup
\fixFloatSize{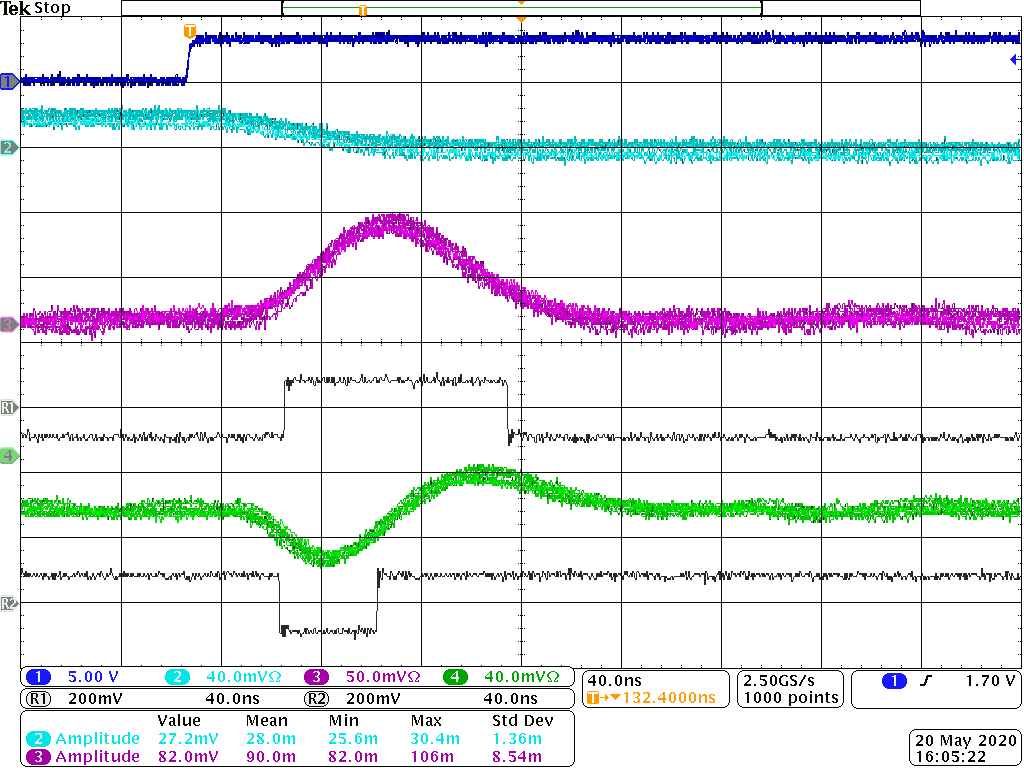}
\begin{figure}[!htbp]
\centering \makeatletter\IfFileExists{analog_monitor2.png}{\includegraphics{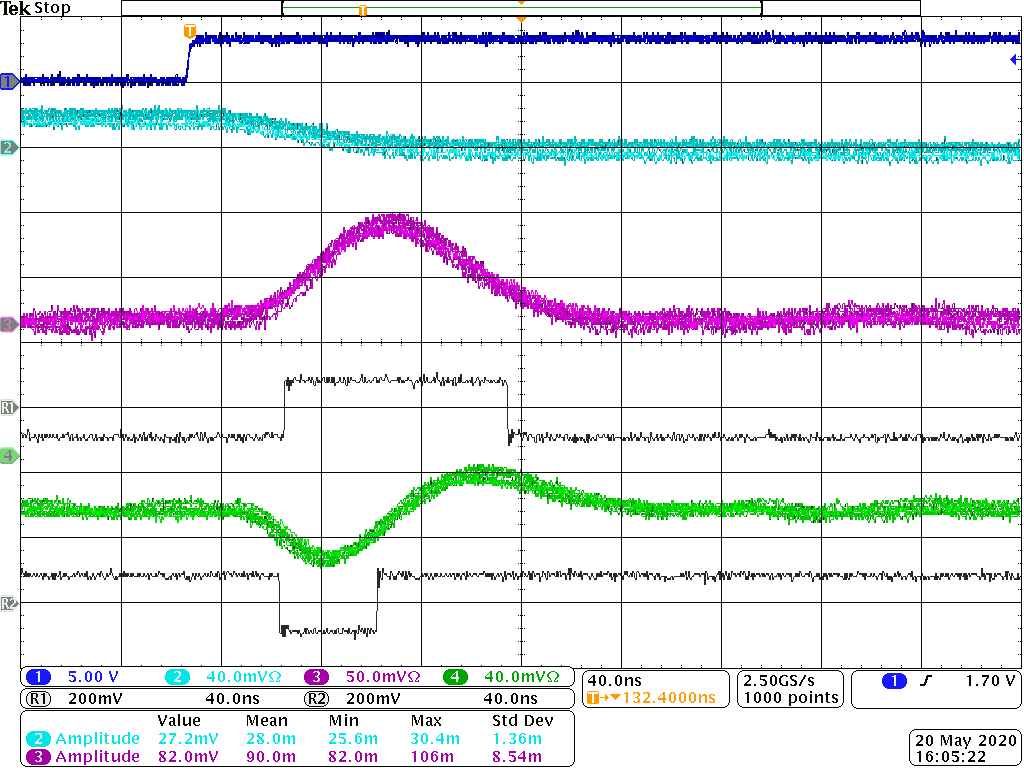}}{}
\makeatother 
\caption{{Output waveform from a typical readout channel. From top to bottom, 
test pulse timing~(blue) and outputs from the CSA~(cyan), CR-RC shaper~(magenta), a comparator for the CR-RC shaper~(black), differentiator~(green), and a 
comparator for the differentiator~(black).
A test pulse injects a 1~MIP signal corresponding to 3.84~fC into the input.}}
\label{fig:analog_monitor}
\end{figure}
\egroup

\bgroup
\fixFloatSize{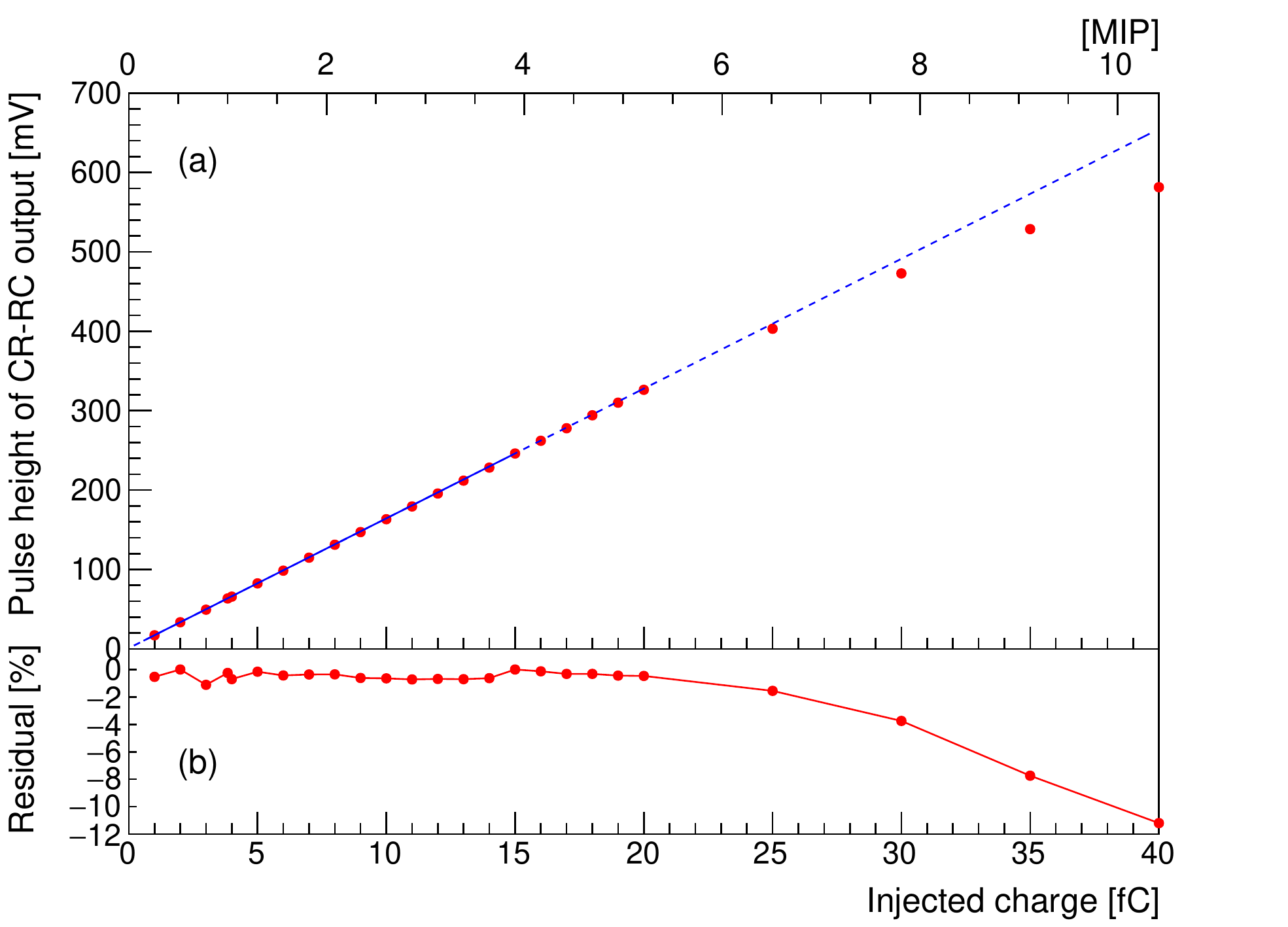}
\begin{figure}[!htbp]
\centering \makeatletter\IfFileExists{pic_paper_dynamic_range.pdf}{\includegraphics{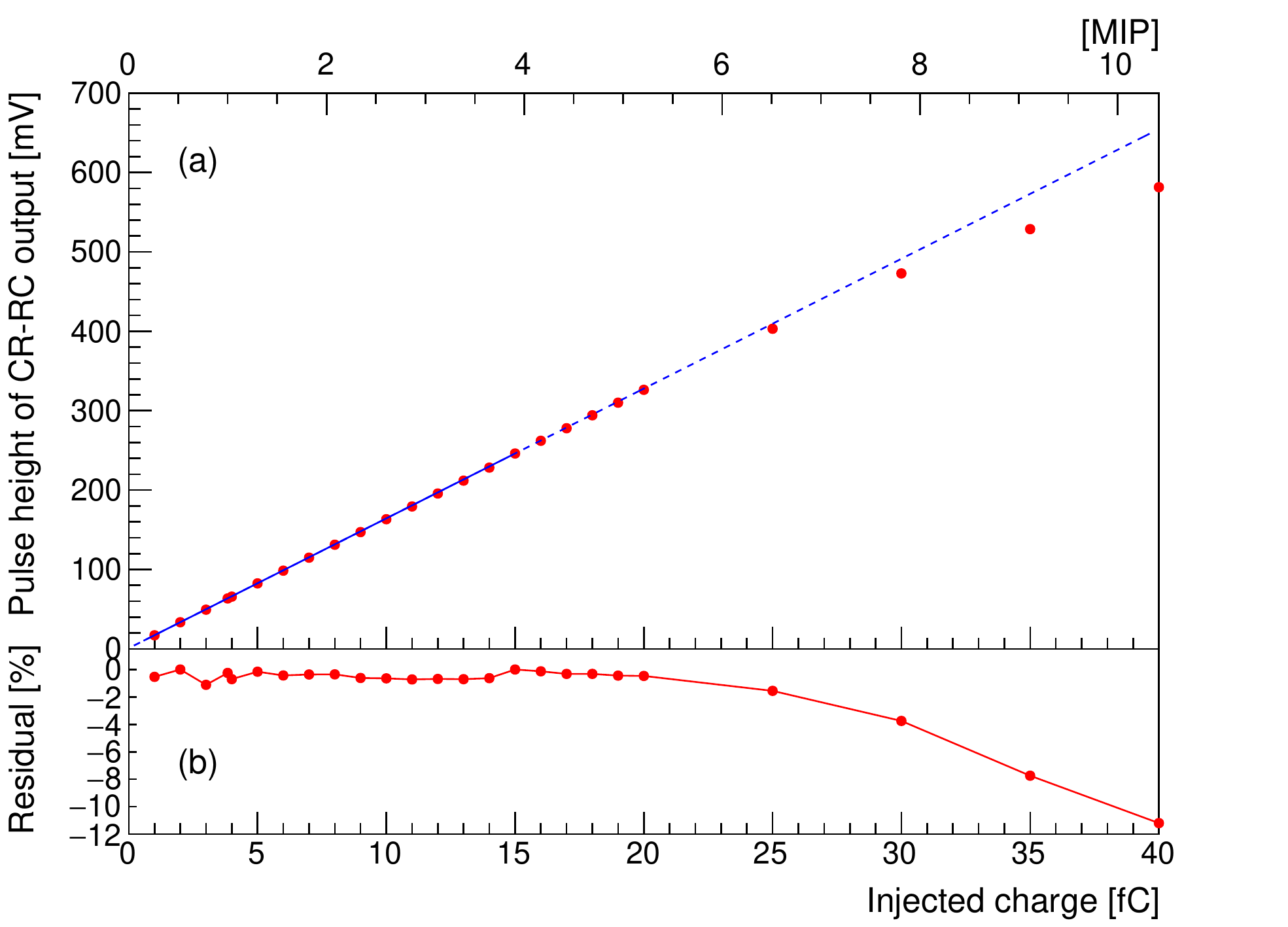}}{}
\makeatother 
\caption{{(a) Pulse height of the CR-RC output and (b) its residual from a straight line as a function of the injected charge.
The straight line is determined from data points at 2~fC and 15~fC.
}}
\label{fig:dynamic_range}
\end{figure}
\egroup

\subsection{S-curve scan and equivalent noise charge}
The noise performance was evaluated by scanning the threshold with different 
injected charges~(s-curve scan).
First, we enabled only the discriminator in the CR-RC filter, and performed 
the threshold scans. 
Fig.~\ref{fig:scurve} shows a counting efficiency as a function of the 
threshold DAC, a so-called s-curve for a typical channel.
The counting efficiency was calculated from the digital outputs.
By fitting a complementary error function, i.e., integrated Gaussian function, 
to the data, the center value and the standard deviation of 
the Gaussian function are obtained.
The center value corresponds to the pulse height.
The standard deviation indicates the noise level.
One least significant bit~(LSB) step size of the threshold DAC for the 
CR-RC shaper is equivalent to the input charge of 0.043~fC.
We repeated this measurement for 32 readout channels, and plot the ENC 
distribution in Fig.~\ref{fig:noise}.
The average equivalent noise charge is measured to be $1547 \pm 75$~electrons 
at $C_{\rm det}=33$~pF,
where the uncertainty indicates the channel-to-channel variation.
If we assume the simulated ENC dependence on the detector capacitance~
(31~electrons/pF), we obtain the ENC to be 1454~electrons at $C_{\rm det}=30$~pF,
which satisfies the requirement for the noise performance.
The simulated ENC is $1210$~electrons at $C_{\rm det}=30$~pF.
The ENC difference between the measurement and the simulation can arise due to various factors:
parasitic input capacitance, digital crosstalk, and noise injection 
from the power supplies.

\bgroup
\fixFloatSize{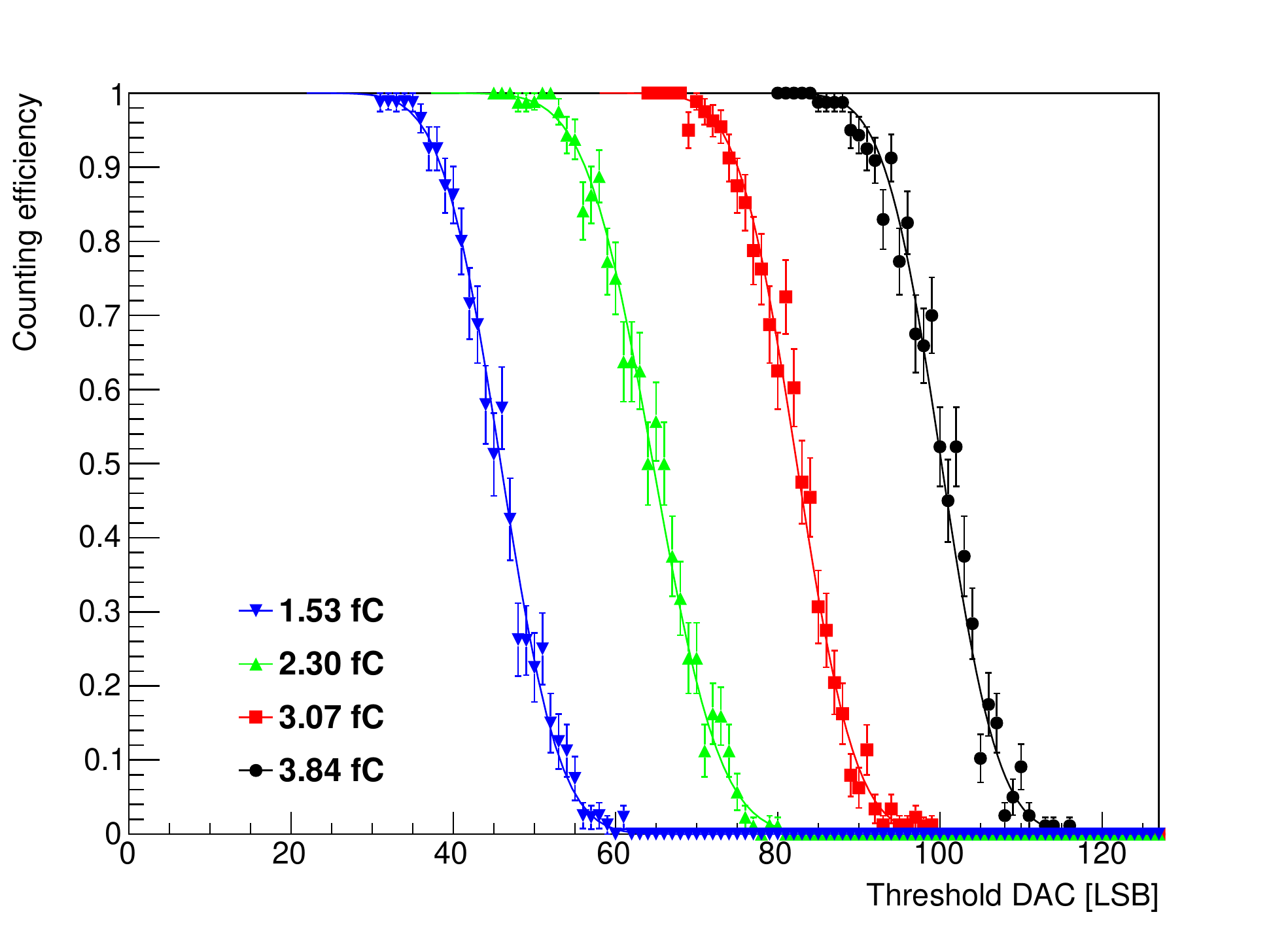}
\begin{figure}[!htbp]
\centering \makeatletter\IfFileExists{pic_paper_scurve.pdf}{\includegraphics{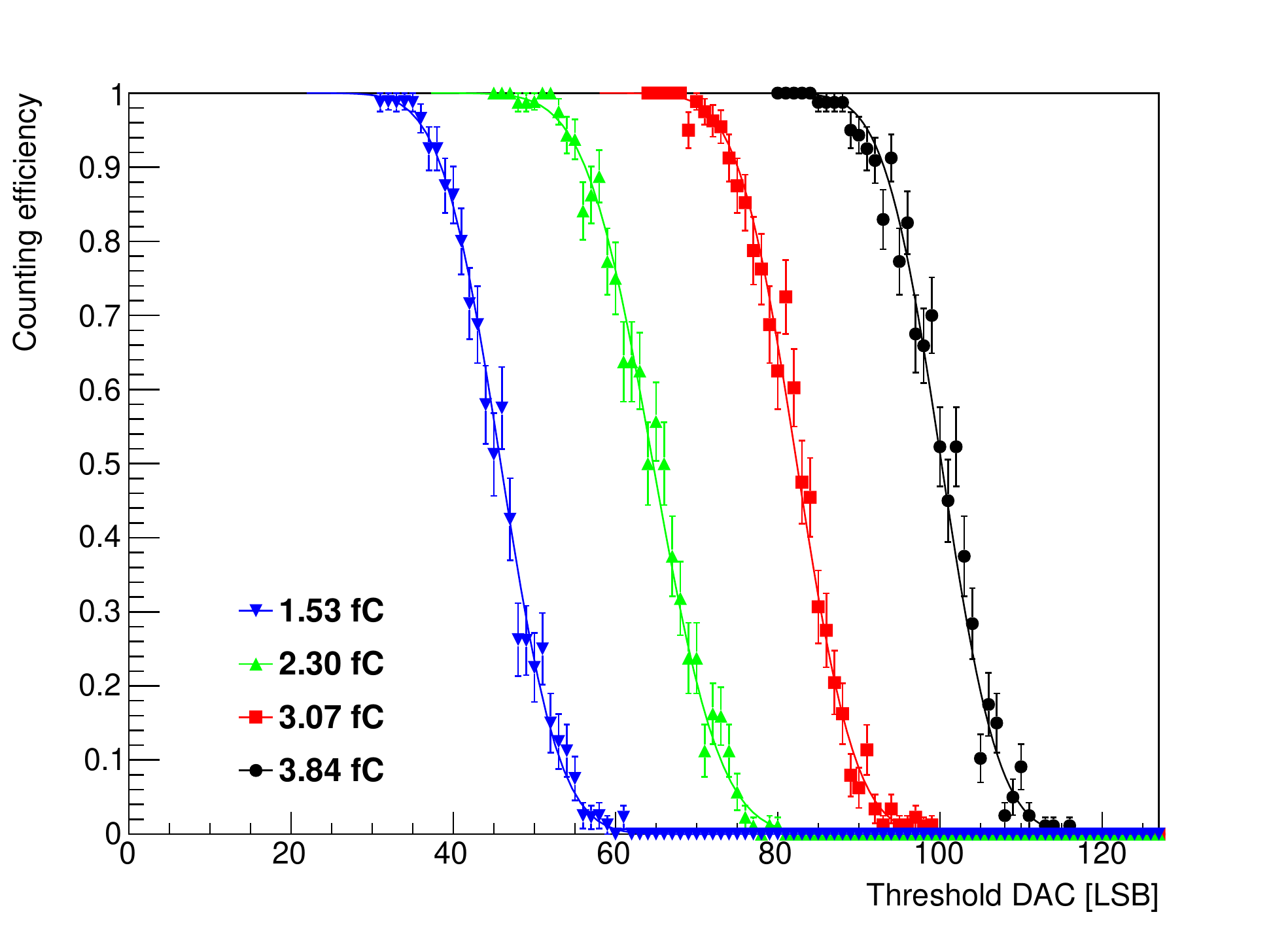}}{}
\makeatother 
\caption{{Counting efficiencies of a typical channel as a function of the threshold DAC in the CR-RC filter~(s-curve). The injected charges are 1.53~fC~(blue), 2.30~fC~(green), 3.07~fC~(red), and 3.84~fC~(black).
A complementary error function is fitted to data.}}
\label{fig:scurve}
\end{figure}
\egroup

\bgroup
\fixFloatSize{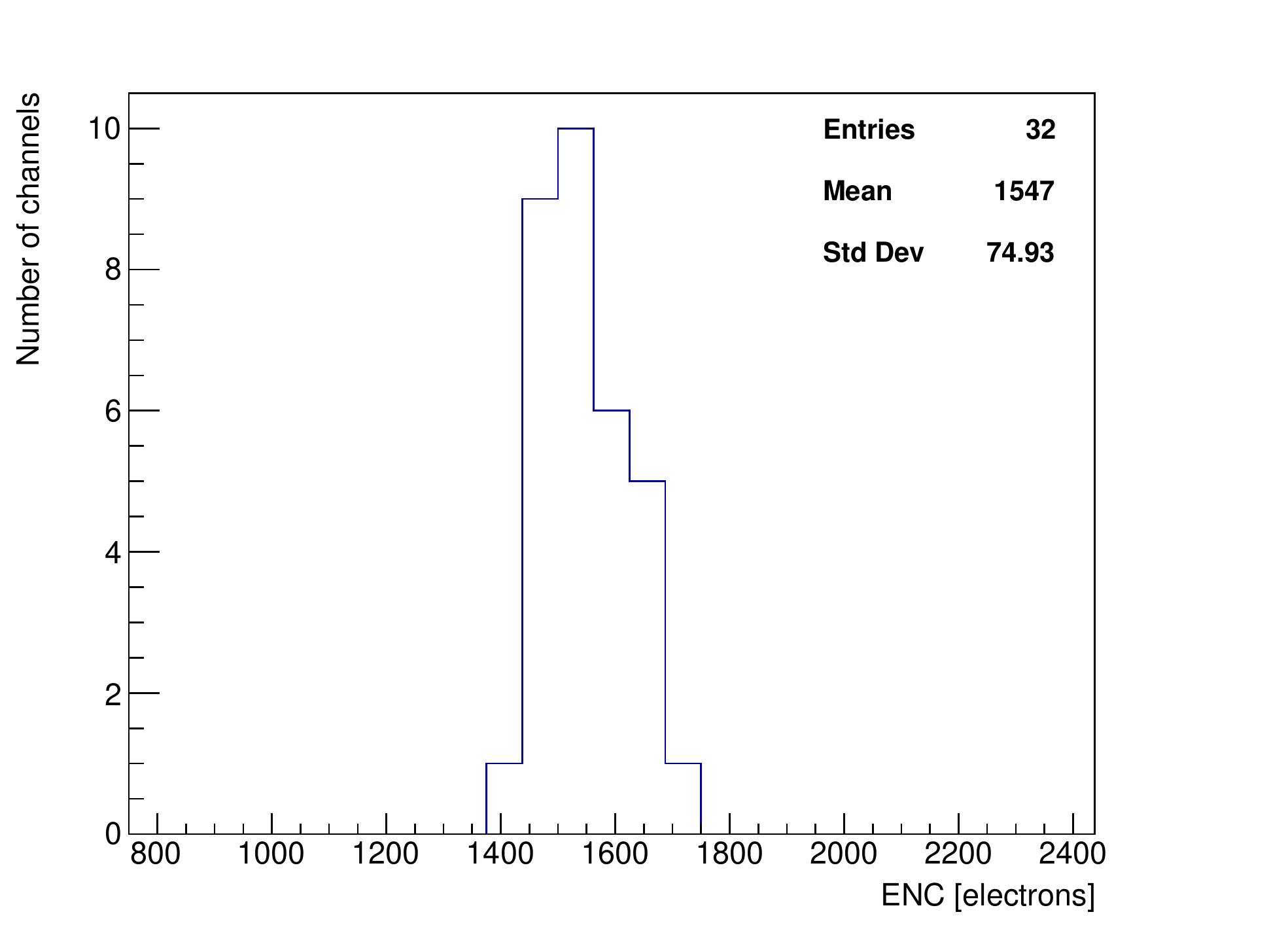}
\begin{figure}[!htbp]
\centering \makeatletter\IfFileExists{pic_paper_noise.pdf}{\includegraphics{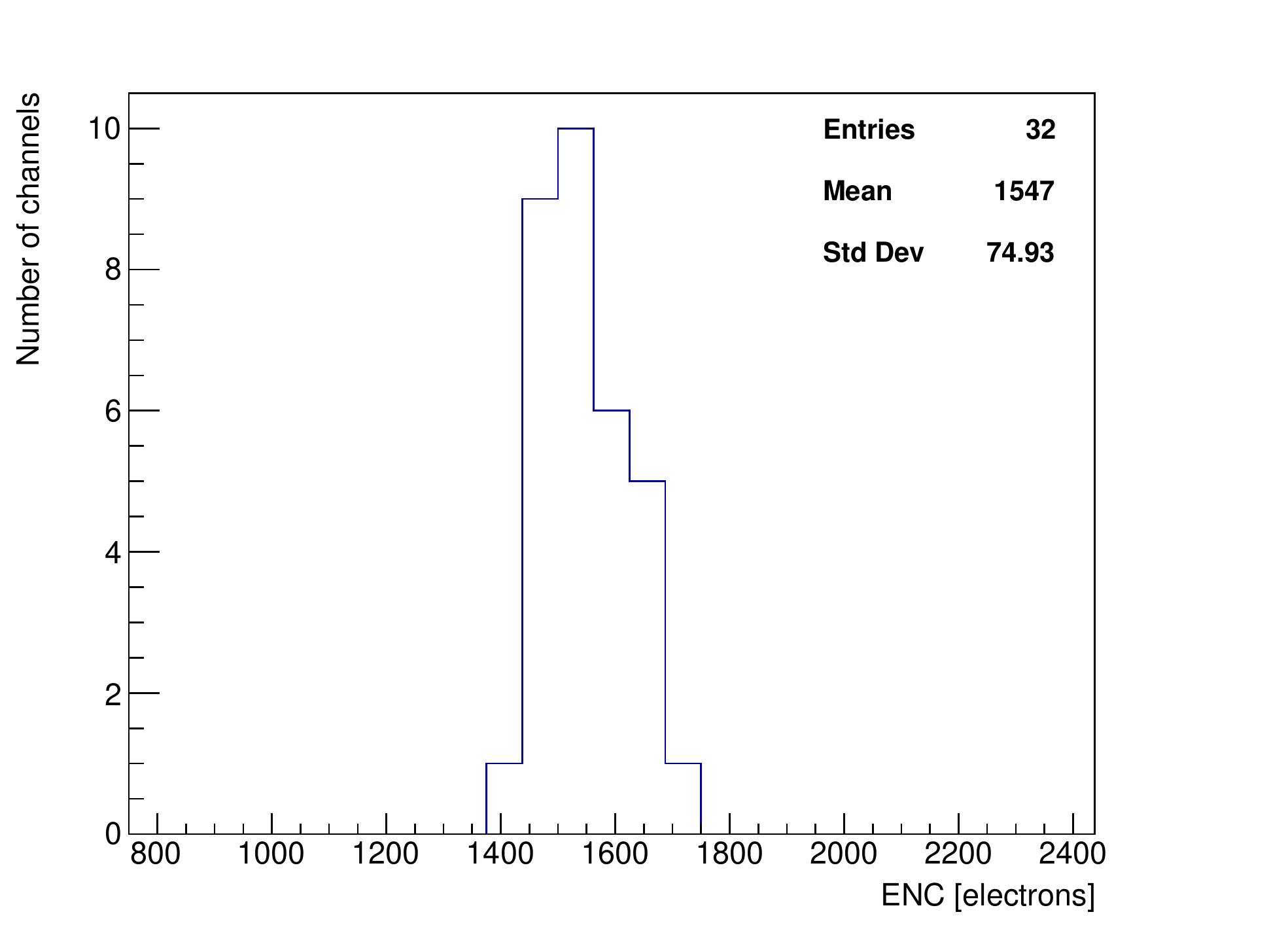}}{}
\makeatother 
\caption{{ENC distribution for 32~readout channels.
}}
\label{fig:noise}
\end{figure}
\egroup

\subsection{Time walk}
Based on the above results, we set the threshold voltage of the CR-RC shaper 
to match 1.53~fC which corresponds to an 0.3~MIP charge level in each channel.
Then the threshold voltage for the voltage differentiator was adjusted to 
minimize the time walk.
Fig.~\ref{fig:timewalk_typical} shows the hit timing vs. threshold voltages 
of the voltage differentiator in a typical readout channel.
The hit timing is defined as the leading-edge timing of the final output, 
which is determined by the differentiator.
A plateau appeared in the hit timing by giving a proper threshold voltage to 
the differentiator.
The time walk of our application is defined as a maximum timing variation in 
the range from 0.5~MIP to 3.0~MIP.\,
Fig.~\ref{fig:timewalk} shows the time walk distribution for the 32~readout 
channels.
This result satisfies the requirement for the time walk
and the mean value of the time walk was $0.38 \pm 0.16$~ns, where the 
uncertainty indicates the channel-to-channel variation.
We note that the hit timing over the injected charges of 3~MIP is increasing, 
however, we confirmed the plateau can be expanded up to 5~MIP by further 
tuning the bias of the shaping amplifiers.

\bgroup
\fixFloatSize{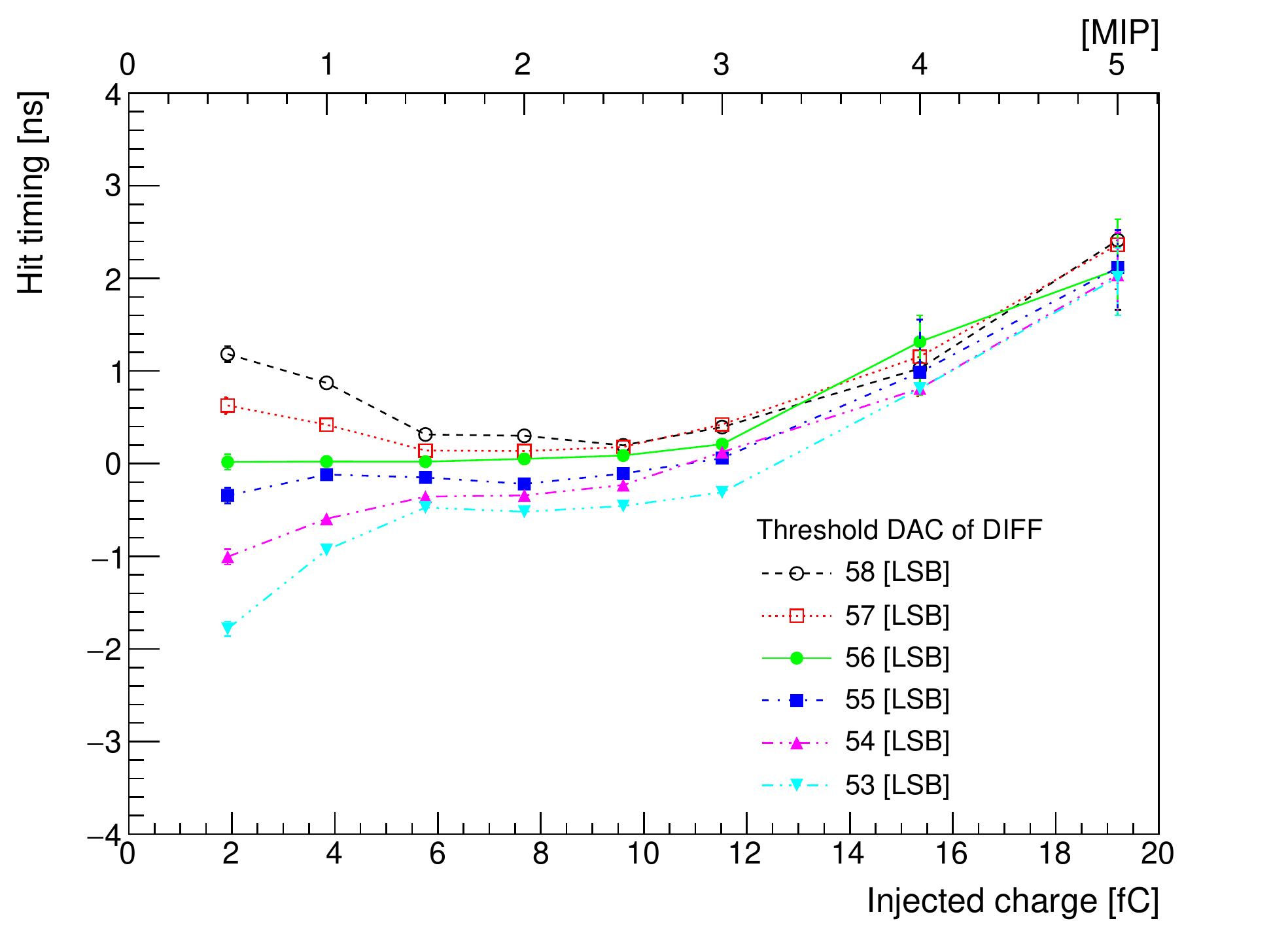}
\begin{figure}[!htbp]
\centering \makeatletter\IfFileExists{pic_paper_timewalk_typical.pdf}{\includegraphics{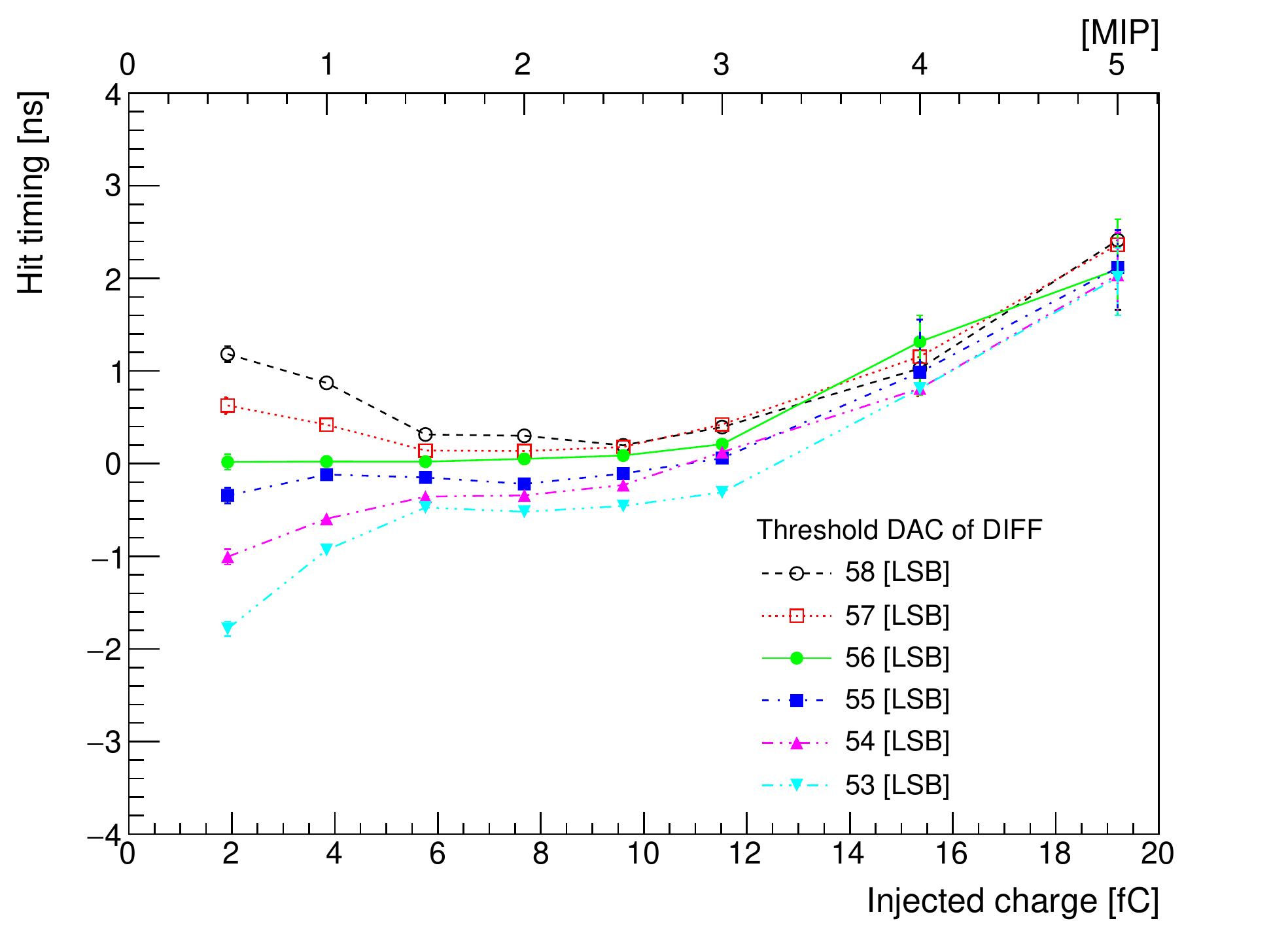}}{}
\makeatother 
\caption{{Hit timing of a typical channel as a function of the injected charges.
Each curve has a different threshold voltage at the differentiator.
The data with a solid green line correspond to an optimum threshold voltage to minimize time walk,
which is defined as a maximum timing variation in the range from 1.92~fC~(0.5 MIP) to 11.52~fC~(3.0~MIP).
Hit timings of all data points are shifted to adjust the data with optimum threshold voltage and the lowest injected charge to $0$~ns.
}}
\label{fig:timewalk_typical}
\end{figure}
\egroup

\bgroup
\fixFloatSize{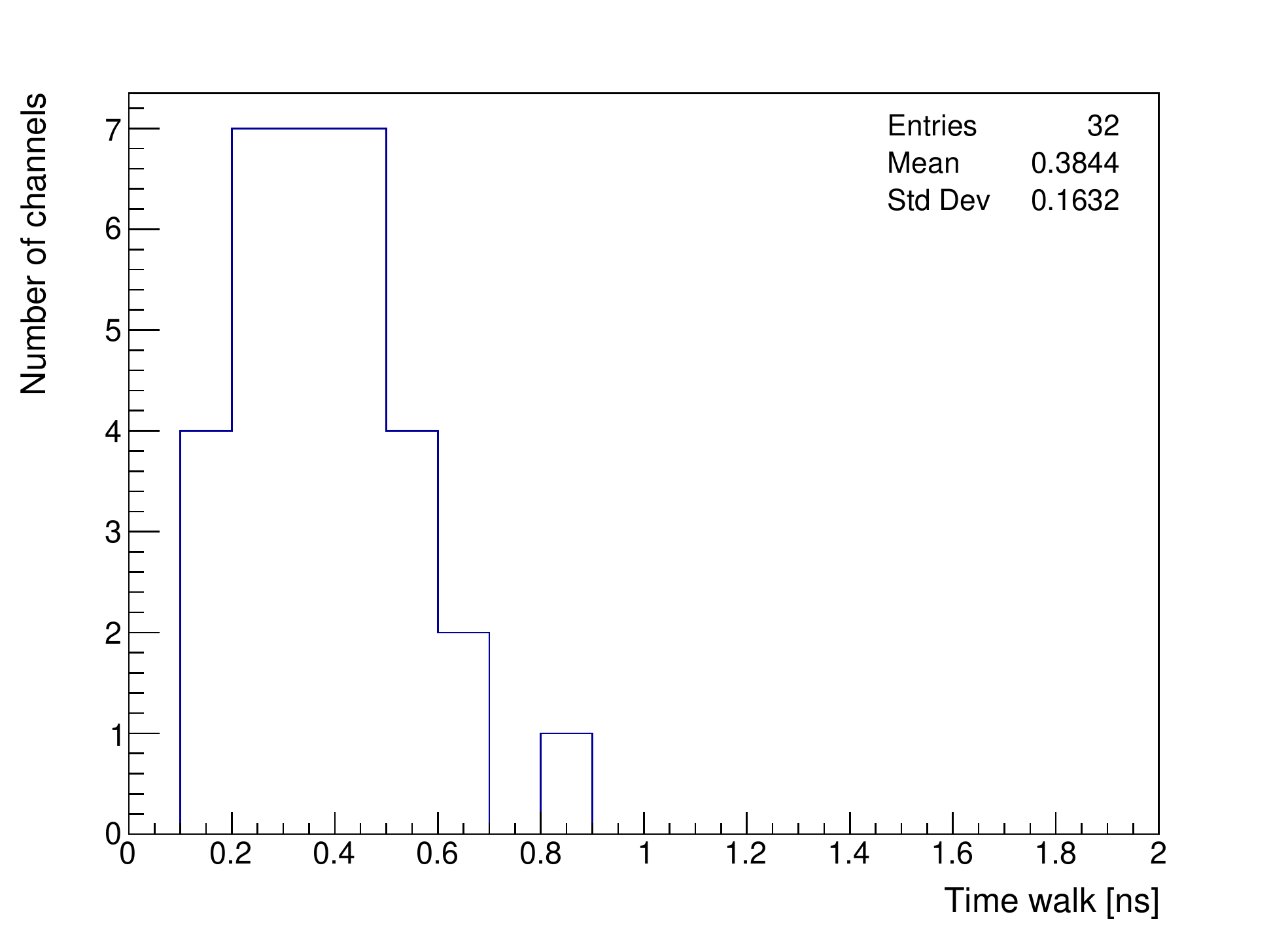}
\begin{figure}[!htbp]
\centering \makeatletter\IfFileExists{pic_paper_timewalk.pdf}{\includegraphics{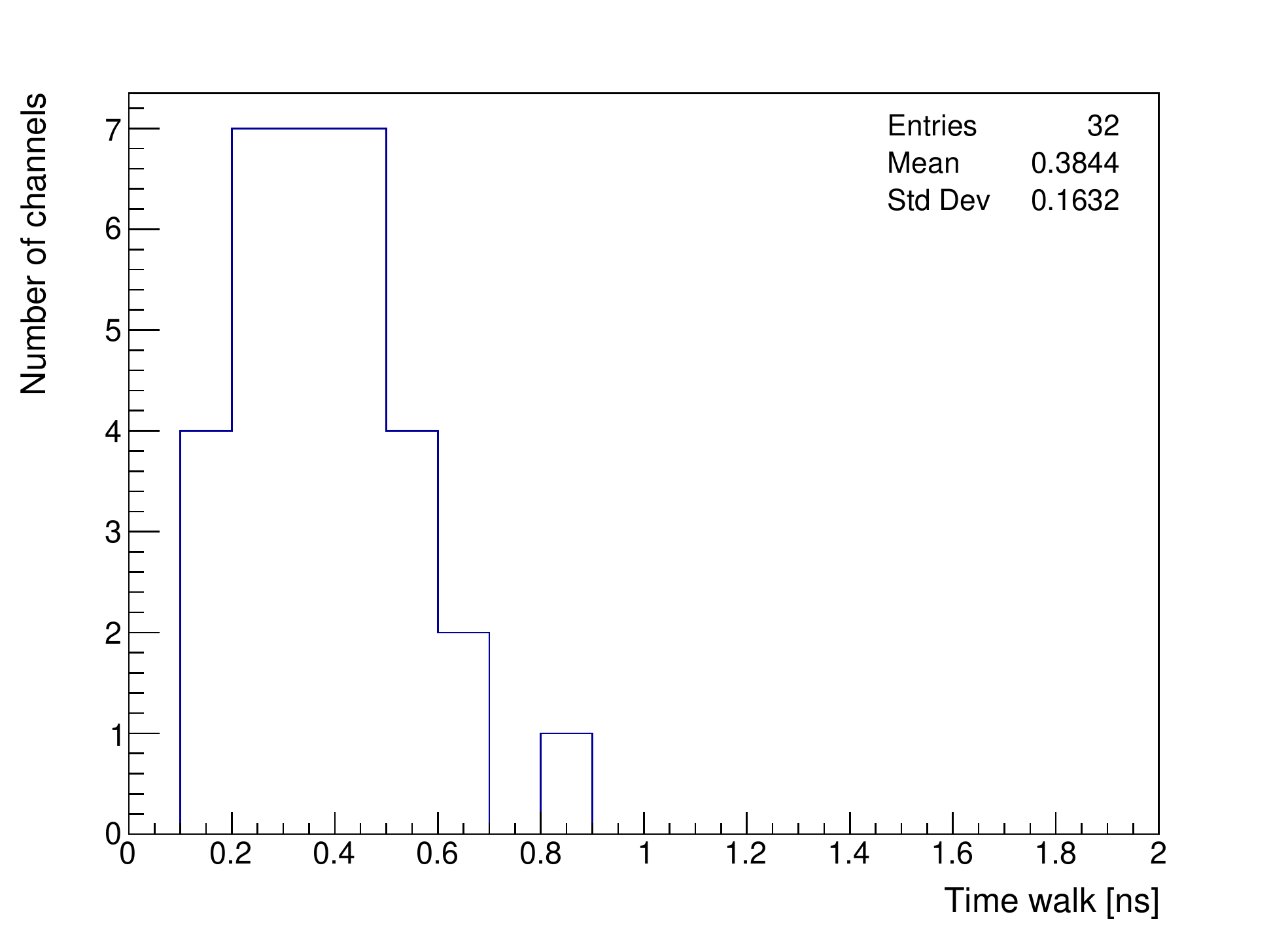}}{}
\makeatother 
\caption{{Time walk distribution for the 32~readout channels.
The time walk is defined as a maximum timing variation in the data set with 
injected charges from 1.92~fC~(0.5 MIP) to 11.52~fC~(3.0~MIP).}}
\label{fig:timewalk}
\end{figure}
\egroup

\subsection{Time over threshold and timing jitter}
The time over threshold~(ToT) as a function of  injected charge is shown 
in Fig.~\ref{fig:tot_and_jitter}(a).
The reconstruction quality of the positron tracks can be improved by 
filtering events based on the ToT information. As described in section~II, 
the leading edge of the hit signal is determined either from the voltage 
differentiator or the single CR-RC shaper by the register setting.
We compared the performance at the normal operating conditions, i.e., using 
outputs from both CR-RC and voltage differentiator in discriminators, in 
the case that the CR-RC path only is used to get the ToT values. Both cases 
are labeled as CRRC\&DIFF (red) and CRRC (blue) in the figure.
Averaged ToT of 32~channels at 1~MIP are determined as $34.1 \pm 0.6$~ns 
(CRRC\&DIFF) and $74.5 \pm 1.6$~ns (CRRC).

We note that the ToT value at 1~MIP of the CRRC case~(74.5~ns) can be 
compared with the simulation value of the pulse width~(75~ns) as shown in 
Table~\ref{tab:requirement}.
The measurement results are close to the designed performance.
The ToT of the CRRC\&DIFF becomes flat above 15~fC.
The non-saturation region can be expanded up to 20~fC by tuning the bias of the shaping amplifiers.
 
 


Fig.~\ref{fig:tot_and_jitter}(b) shows the timing jitter~($\sigma$) 
as a function of the injected charge.
The timing jitter becomes worse for smaller signals, since the leading edge 
of the hit signal is directly affected by the electrical noise.
The average timing jitter at 1~MIP for 32~channels is $2.9 \pm 0.1$~ns for 
the CRRC\& DIFF case and $2.5 \pm 0.1$~ns for the CRRC case.
The jitters at 0.5~MIP is $4.8 \pm 0.2$~ns for the CRRC \& DIFF, and 
$4.5 \pm 0.2$~ns for the CRRC.
They are comparable with the sampling interval of 5~ns.

\bgroup
\fixFloatSize{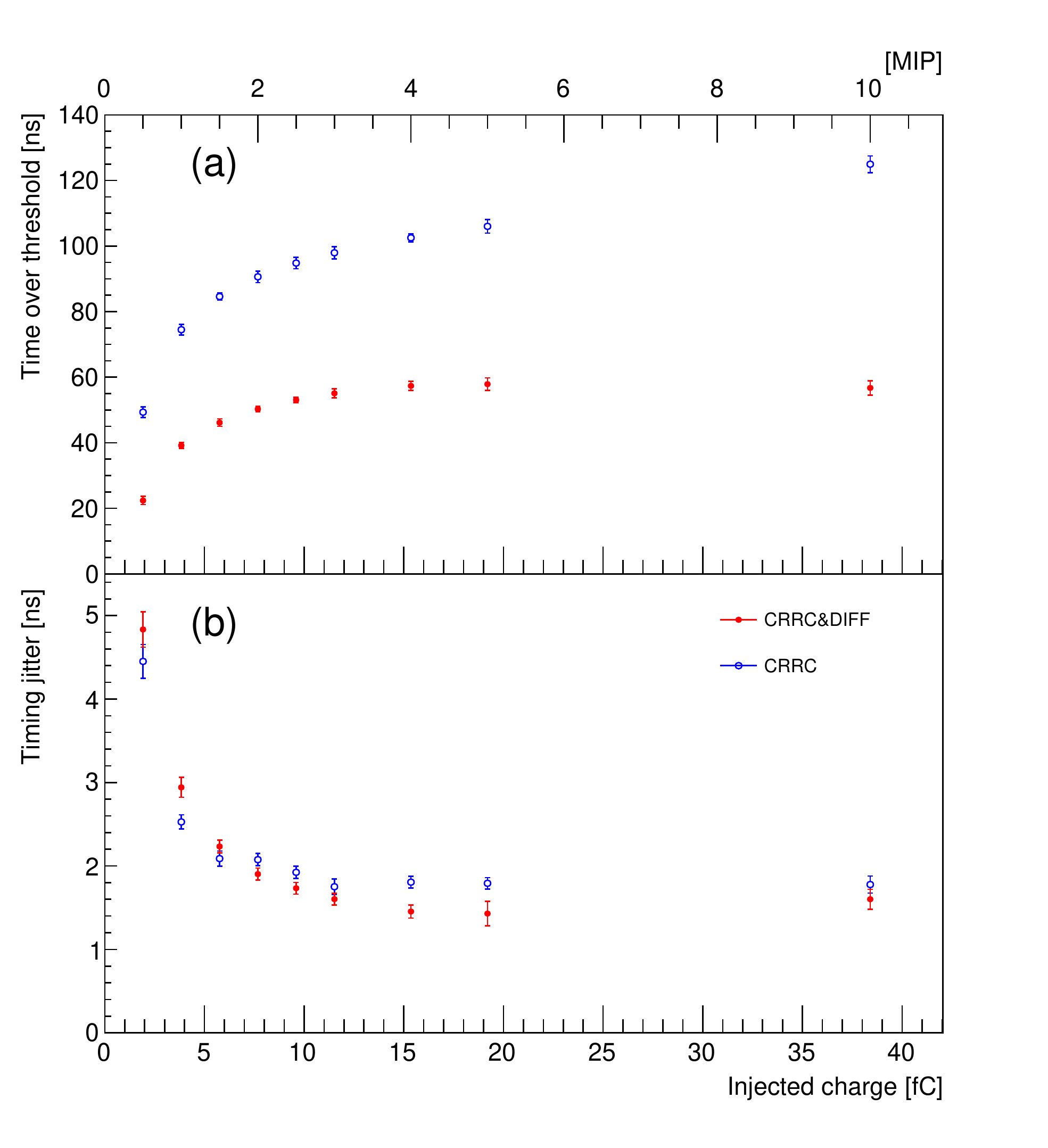}
\begin{figure}[!htbp]
\centering \makeatletter\IfFileExists{pic_paper_weighted_tot_and_jitter.pdf}{\includegraphics{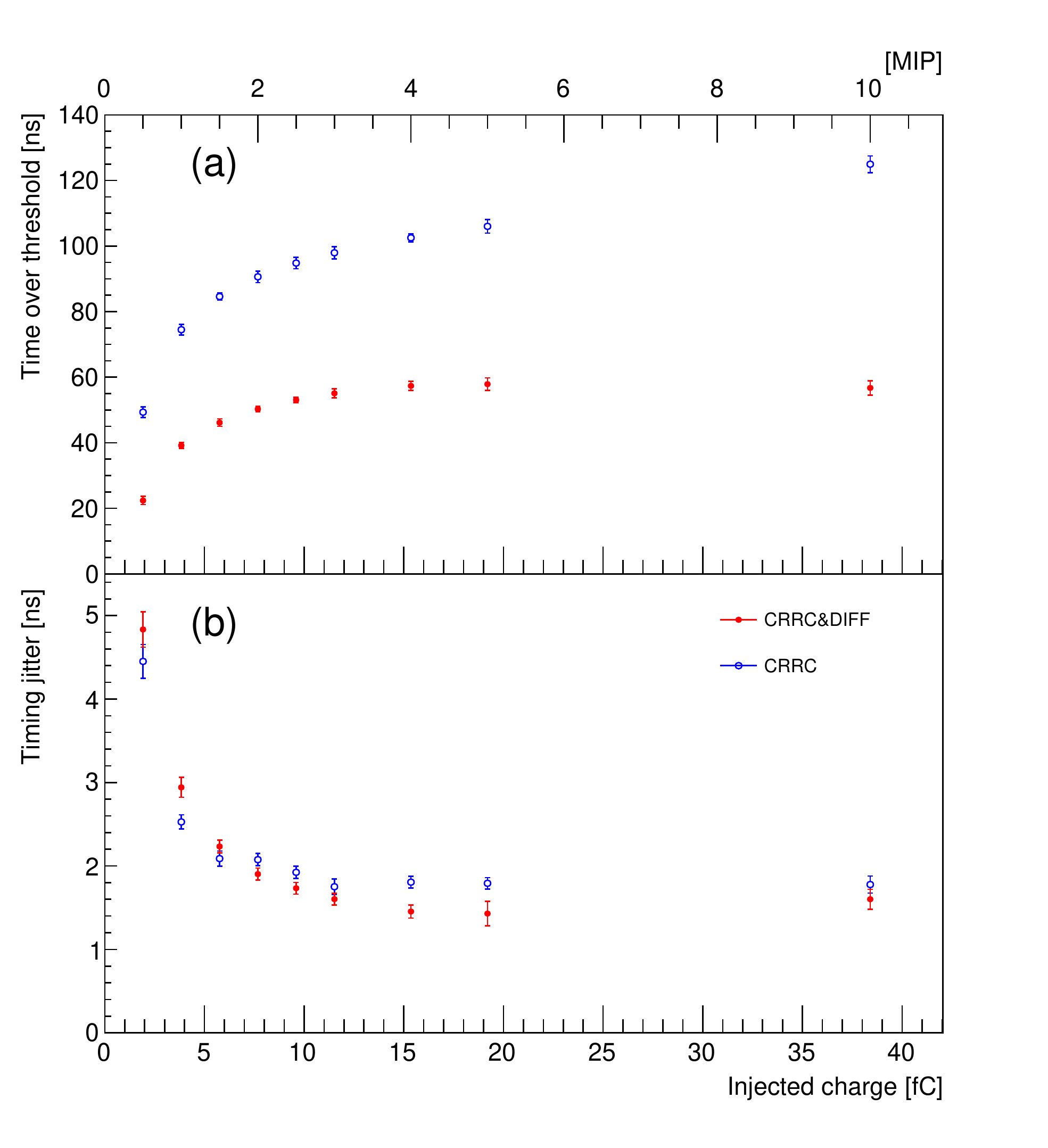}}{}
\makeatother 
\caption{{(a) Time over threshold and (b) timing jitter as a function of the injected charge for the 32~readout channels.
The result with normal operation, in which both comparators for the CR-RC 
shaper and the differentiator are enabled, is shown by solid red circles.
The result with operation, in which a comparator is enabled for the CR-RC 
shaper only, is shown by open blue circles.
The threshold voltage of the CR-RC shaper was set to be at the 0.3~MIP 
charge level.
The threshold voltage of the differenetiator was adjusted to minimize 
the time walk.
The error bars indicate the channel variation.}}
\label{fig:tot_and_jitter}
\end{figure}
\egroup

\begin{table*}[!htbp]
\begin{threeparttable}
\caption{{Summary of the SliT128C performance} }
\label{tab:requirement}
\def\arraystretch{1}
\ignorespaces 
\centering 
\begin{tabulary}{\linewidth}{p{\dimexpr.2933\linewidth-2\tabcolsep}p{\dimexpr.2067\linewidth-2\tabcolsep}p{\dimexpr.25\linewidth-2\tabcolsep}p{\dimexpr.25\linewidth-2\tabcolsep}}
\hline  & Requirement & Simulation & Measurement\\
\hline 
Peaking time &
$<75$~ns   & 35.1~ns
&
64.2~ns\tnote{a}
  \\
Pulse width at 1~MIP&
$<100$~ns  & 75.0~ns
&
74.5~ns   
  \\
Dynamic range &
$>4$~MIP   & 8~MIP  & $>7.8$~MIP
  \\
ENC &
$<1,600$~$e^-$@$C_{\rm det}=30$~pF  &
1210~$e^-$@$C_{\rm det}=30$~pF &
$1547 \pm 75$~$e^-$@$C_{\rm det}=33$~pF
\\
Time walk~(0.5-3.0~MIP) &
$<1$~ns   & 0.4~ns
   &
$0.38 \pm 0.16$~ns
  \\
Jitter at 0.5~MIP &
$<5$~ns   & 4.89~ns
   & $4.8 \pm 0.2$~ns
  \\
Power consumption &
0.64~W/chip   &
N.A.   &
0.30~W/chip
  \\
\hline 
\end{tabulary}\par 
\begin{tablenotes}
\item[a] Outputs measured from the monitor which include delays due to the 
parasitic capacitance and finite buffer drive strength.
\end{tablenotes}
\end{threeparttable}
\end{table*}

When the number of fired channels are increased, the instantaneous IR drop becomes a main issue to degrade the timing performance. This issue is related with the implementation of decoupling capacitors between supply powers and ground. In our application, the fired channels are estimated to be a few dozen channels at most, and thus, the performance degradation can be negligible in our application.

\subsection{Power consumption}
We supplied 1.8~V for an analog circuit and 1.8~V for a digital one.
An additional dedicated bias voltage of 1.1~V is applied to the input 
transistors in the CSA for noise optimization and avoiding IR-drops.
By measuring the current at each power line, power consumption was 
determined to be 0.13~W, 0.11~W, and 0.06~W for the analog, the digital, and 
the CSA, respectively.
The total power consumption is 0.30~W per chip.

\section{Conclusions}
We have developed a series of silicon-strip readout chips named  SliT 
with the 180~nm CMOS technology.
The SliT is designed for the precise measurement of the muon  $g-2$/EDM, which 
is planned at J-PARC in Japan. The main objective of the SliT is to provide 
the timing information for positron tracks from silicon-strip sensors, with 
amplitude-independent time walks.
We have also developed a new readout chip named ``SliT128C''.
In this chip, the leading edge is generated from the timing of 
baseline-crossing by the bipolar waveforms. This zero-crossing architecture is 
implemented by adding the voltage differentiator at the output of the  
CR-RC shaper. As a result, the time walk effect is suppressed to 
subnanosecond~($0.38\pm0.16$~ns).
The equivalent noise charge is $1547 \pm 75~e^{\rm -}$ with $C_{\rm det}=33$~pF. 
The SliT128C consists of 128~channels, which are completely functional.
Other evaluated performance results are also summarized in Table~\ref{tab:requirement}.
The SliT128C satisfies all requirements of the J-PARC muon $g-2$/EDM experiment.


\section*{Acknowledgment}
The authors would like to thank O.~Sasaki, H.~Ikeda, and M. Miyahara for constructive 
comments and useful discussions concerning the design of SliT.
The authors also wish to thank the Open Source Consortium of 
Instrumentation~(Open-It) of KEK.
This work was supported by the JSPS KAKENHI (Grant No. JP15H05742 and 
JP18H05226) and the 37th CASIO research funding program (Grant No. 26).



%



\end{document}